\pdfoutput=1
\documentclass[cernpreprint, atlasdraft=false, UKenglish, texmf, orcidlogo]{atlasdoc}
 
\usepackage{atlaspackage}
\usepackage{atlasbiblatex}
 
\usepackage[process, jetetmiss]{atlasphysics}
 
\graphicspath{{logos/}{figures/}}
 
\usepackage{ANA-EXOT-2018-34-PAPER-defs}
\AtlasTitle{Search for doubly charged Higgs boson production in multi-lepton final states using 139~fb$^{-1}$ of proton--proton collisions at $\sqrt{s}$ = 13~TeV with the ATLAS detector}

\AtlasAbstract{A search for pair production of doubly charged Higgs bosons ($H^{\pm \pm}$), each decaying into a pair of prompt, isolated, and highly energetic leptons with the same electric charge, is presented. The search uses a proton--proton collision data sample at a centre-of-mass energy of 13~TeV corresponding to an integrated luminosity of 139~fb$^{-1}$ recorded by the ATLAS detector during Run~2 of the Large Hadron Collider (LHC). This analysis focuses on same-charge leptonic decays, $H^{\pm \pm} \rightarrow \ell^{\pm} \ell^{\prime \pm}$ where $\ell, \ell^\prime=e, \mu, \tau$, in two-, three-, and four-lepton channels, but only considers final states which include electrons or muons. No evidence of a signal is observed. Corresponding upper limits on the production cross-section of a doubly charged Higgs boson are derived, as a function of its mass $m(H^{\pm \pm})$, at 95\% confidence level. Assuming that the branching ratios to each of the possible leptonic final states are equal, $\mathcal{B}(H^{\pm \pm} \rightarrow e^\pm e^\pm) = \mathcal{B}(H^{\pm \pm} \rightarrow e^\pm \mu^\pm) = \mathcal{B}(H^{\pm \pm} \rightarrow \mu^\pm \mu^\pm) = \mathcal{B}(H^{\pm \pm} \rightarrow e^\pm \tau^\pm) = \mathcal{B}(H^{\pm \pm} \rightarrow \mu^\pm \tau^\pm) = \mathcal{B}(H^{\pm \pm} \rightarrow \tau^\pm \tau^\pm) = 1/6$, the observed (expected) lower limit on the mass of a doubly charged Higgs boson is 1080~GeV (1065~GeV) within the left-right symmetric type-II seesaw model, which is the strongest limit to date produced by the ATLAS Collaboration. Additionally, this paper provides the first direct test of the Zee--Babu neutrino mass model at the LHC, yielding an observed (expected) lower limit of $m(H^{\pm \pm})$ = 900~GeV (880~GeV).}

\author{The ATLAS Collaboration}
 
\AtlasRefCode{EXOT-2018-34}
 
\PreprintIdNumber{CERN-EP-2022-212}

\arXivId{2211.07505}
 
\HepDataRecord{138987}
 
\AtlasJournal{\EPJC}
\AtlasJournalRef{\EPJC 83 (2023) 605}
\AtlasDOI{10.1140/epjc/s10052-023-11578-9}

\hypersetup{pdftitle={ATLAS document},pdfauthor={The ATLAS Collaboration}}
 
\begin{document}
 
\maketitle
 
\tableofcontents

\clearpage
\section{Introduction}

In the Standard Model (SM), no doubly charged bosons are present. However, various theories beyond the Standard Model (BSM theories), namely type-II seesaw models~\cite{Cai:2017mow,muhlleitner2003note,akeroyd2005single,hektor2007testing,perez2008testing,chao2008correlative}, left-right symmetric models (LRSMs)~\cite{Pati:1974yy,Mohapatra:1974hk,Senjanovic:1975rk,BhupalDev:2016nfr,Borah:2016hqn,Senjanovic:2016bya}, the Zee--Babu neutrino mass model~\cite{Zee:1985rj,Zee:1985id,Babu:1988ki}, 3-3-1 models~\cite{Corcella:2018eib}, and the Georgi--Machacek model~\cite{Georgi:1985nv}, predict the existence of doubly charged bosons decaying into same-charge lepton pairs. In left-right symmetric models based on the $SU(2)_L \times SU(2)_R \times U(1)_{B-L}$ symmetry, Higgs multiplets $\Delta_L$ and $\Delta_R$ transforming, respectively, as triplets under $SU(2)_L$ and $SU(2)_R$ gauge symmetries, contain doubly charged Higgs bosons, termed $\DCH_L$ and $\DCH_R$. Doubly charged Higgs particles can therefore couple to either left-handed or right-handed leptons. The cross-section for the \DY production of $H^{++}_L H^{--}_L$ is a factor of about two larger than for the $H^{++}_R H^{--}_R$ production, due to the different couplings to the \Zboson boson~\cite{Fuks:2019clu}. Since the $\DCH_L$ particle is common to LRSMs and the canonical type-II seesaw model, the results can be directly interpreted in both. In the Zee--Babu case, two complex scalar $SU(2)_L$ singlets are proposed within the SM gauge group, where one of them is doubly charged and is usually denoted by $k^{\pm \pm}$. As it has the same quantum numbers as the $\DCH_R$ from LRSMs, their electroweak production is the same.\footnote{In principle, the $\DCH_R$ also couples to a $\Zboson^{\prime}$ boson, which makes their production cross-sections distinguishable. However, the $\Zboson^{\prime}$ boson is assumed to have a mass that is well beyond the energy reach of the LHC, thus making its effects negligible.} It was recently shown that for the \DY production mechanism studied in this analysis, cross-sections and differential scalar distributions in the Zee--Babu and type-II seesaw models differ at most by a normalisation factor when all theoretical inputs are the same~\cite{Ruiz:2022sct}. If not stated explicitly, \DCH represents any of the $\DCH_L$, $\DCH_R$, or $k^{\pm \pm}$ particles throughout the paper.
 
\begin{figure}[htbp]
\centering
\includegraphics[width=0.45\textwidth]{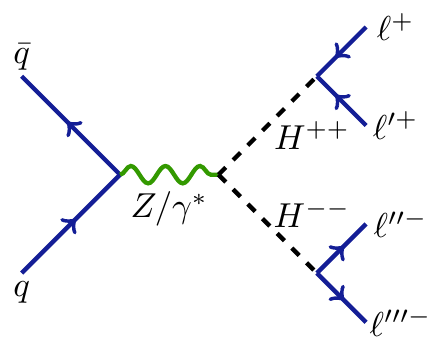}
 
\caption{Feynman diagram of the pair production process $\pp \rightarrow H^{++} H^{--}$. While the analysis allows for \DCH decays into all lepton flavour combinations, it only studies electrons and muons in the final states.}
\label{fig:DCHfey}
\end{figure}

The Feynman diagram of the \DY pair production mechanism considered in this analysis is shown in \Fig{\ref{fig:DCHfey}}.
 
At the LHC, other mechanisms such as vector-boson fusion, gluon--gluon fusion, and photon-initiated~\cite{Babu:2016rcr} processes are less important than \DY production~\cite{Fuks:2019clu} in the mass range of interest for this paper, which is \SI{400}{\GeV} to \SI{1300}{\GeV}.
 
The ATLAS Collaboration previously analysed data corresponding to an integrated luminosity of \SI{36.1}{\ifb} from $\sqrt{s}=\SI{13}{\TeV}$ \pp collisions at the LHC during the 2015 and 2016 data-taking periods. Masses of doubly charged Higgs bosons were excluded up to \SI{870}{\GeV} for $\DCH_L$ and up to \SI{760}{\GeV} for $\DCH_R$ at \SI{95}{\%} confidence level (\CL)~\cite{EXOT-2016-07}. The CMS Collaboration performed a similar search using $\sqrt{s}=\SI{7}{\TeV}$ \pp collisions collected during \RunOne~\cite{CMS-HIG-12-005}, establishing lower bounds on the \DCH mass between \SI{204}{\GeV} and \SI{459}{\GeV} in the \SI{100}{\%} branching fraction scenarios, depending on the flavour content of the two-lepton final states. A complementary analysis from the ATLAS Collaboration searched for doubly charged Higgs bosons in the $\DCH \rightarrow W^\pm W^\pm$ decay channel using the whole $\SI{139}{\ifb}$ dataset from \RunTwo of the LHC and excluded \DCH masses up to \SI{350}{\GeV}~\cite{HDBS-2019-06}.
 
The doubly charged Higgs boson can decay to $\Wboson \Wboson$ or $\ell \ell^\prime$ ($\ell, \ell^\prime=e,~\mu,~\tau$), depending on the vacuum expectation value of the $\Delta_L$ Higgs triplet, $v_{\Delta}$~\cite{perez2008testing}. In this paper, it is assumed that $v_{\Delta}$ is smaller than \SI{e-8}{\GeV}. Therefore, only leptonic decays of doubly charged Higgs bosons, $\DCH \rightarrow \ell^{\pm} \ell^{\prime \pm}$, are relevant.
 
A measurement of same-charge lepton pairs is sensitive to the same-charge partial decay width of \DCH given by:
 
\begin{equation}
\Gamma (\DCH \rightarrow \ell^{\pm} \ell^{\prime \pm}) = k\frac{h^{2}_{\ell \ell^\prime}}{16 \pi} m(\DCH),
\label{eq:decay-rate}
\end{equation}
 
with $k=2$ if the two leptons have the same flavour and $k=1$ if their flavours are different. The factor $h_{\ell \ell^\prime}$ has an upper bound that depends on the flavour combination~\cite{Dutta:2014dba,Rentala:2011mr} but it is set to an allowed value $h_{\ell \ell^\prime} = 0.02$ for all flavour combinations in this analysis. This choice corresponds to prompt decays of the \DCH bosons and, according to \Eqn{(\ref{eq:decay-rate})}, results in a mean proper decay length $c\tau \sim \SI{10}{\femto\metre}$ across the whole \DCH mass range considered in this search. In general, there is no preference for decays into heavier leptons, as the coupling is not proportional to the lepton mass as it is for the SM Higgs boson. Therefore, the couplings of \DCH to all three SM charged leptons ($e,~\mu,~\tau$) were assumed to be the same.
 
It is worth stressing that lepton-flavour-violating decays such as $\DCH \rightarrow e^\pm \mu^\pm$, $\DCH \rightarrow \mu^\pm \tau^\pm$ or $\DCH \rightarrow e^\pm \tau^\pm$ are also allowed by the model. This fact connects high-energy searches at LHC with low-energy neutrinoless double beta decay searches~\cite{Tello:2010am}.
 
The analysis presented in this paper searches for leptonic decays $\DCH \rightarrow \ell^\pm \ell^{\prime \pm}$ of doubly charged Higgs boson pairs, resulting in same-charge lepton pairs in final states with two, three or four leptons. Since these events are produced very rarely in proton--proton (\pp) collisions by SM processes, same-charge lepton pairs represent a striking signature for BSM physics. The analysis only considers light leptons ($e$ and $\mu$) in the two-, three- and four-lepton final states, thereby including leptonic $\tau$ decays, using the whole \SI{139}{\ifb} dataset from \RunTwo of the LHC. The assumption that the branching ratios to each of the possible leptonic final states are equal, $\BR(\DCH \rightarrow e^\pm e^\pm) = \BR(\DCH \rightarrow e^\pm \mu^\pm) = \BR(\DCH \rightarrow \mu^\pm \mu^\pm) = \BR(\DCH \rightarrow e^\pm \tau^\pm) = \BR(\DCH \rightarrow \mu^\pm \tau^\pm) = \BR(\DCH \rightarrow \tau^\pm \tau^\pm) = 1/6$, is made. To estimate the compatibility of the data with the SM expectations, as well as to extract upper limits on the \DCH mass at \SI{95}{\%} \CL, a binned maximum-likelihood fit to data is performed in all control and signal regions, described in \Sect{\ref{sec:eventSelection}}. A simultaneous fit to the distribution of the invariant mass of the two same-charge leptons with the highest \pt in the event is implemented in the two- and three-lepton regions, while a single-bin event yield is used in the four-lepton regions, merging two-lepton channels to obtain the best sensitivity.
 
This paper is organised as follows. The ATLAS detector is described in \Sect{\ref{sec:detector}}, the data and simulated events used in the analysis are outlined in \Sect{\ref{sec:monteCarloProduction}}, and the event reconstruction procedure and the analysis strategy are detailed in \Sect{\ref{sec:eventSelection}}. The background estimation is presented in \Sect{\ref{sec:backgroundEstimation}}, and the systematic uncertainties are described in \Sect{\ref{sec:systematics}}. Finally, results and their statistical interpretation are presented in \Sect{\ref{sec:results}}, followed by the conclusions in \Sect{\ref{sec:conclusion}}.
 
\FloatBarrier

\clearpage

\clearpage
\section{ATLAS detector}
\label{sec:detector}
 
\newcommand{\AtlasCoordFootnote}{
ATLAS uses a right-handed coordinate system with its origin at the nominal interaction point (IP) in the centre of the detector and the \(z\)-axis along the beam pipe. The \(x\)-axis points from the IP to the centre of the LHC ring, and the \(y\)-axis points upwards. Cylindrical coordinates \((r,\phi)\) are used in the transverse plane, \(\phi\) being the azimuthal angle around the \(z\)-axis. The pseudorapidity is defined in terms of the polar angle \(\theta\) as \(\eta = -\ln \tan(\theta/2)\). Angular distance is measured in units of \(\Delta R \equiv \sqrt{(\Delta\eta)^{2} + (\Delta\phi)^{2}}\).
}
 
The ATLAS detector~\cite{PERF-2007-01} is a multipurpose particle detector covering nearly $4\pi$ in solid angle in one of the \pp interaction regions of the LHC. It consists of several subdetectors assembled with a cylindrical symmetry coaxial with the beam axis.\footnote{\AtlasCoordFootnote} The inner detector (ID) surrounding the beam pipe is immersed in a \SI{2}{\tesla} solenoidal magnetic field and provides charged-particle tracking in the range $\abseta < 2.5$. Starting from the beam pipe and going outwards, the ID is composed of a high-granularity silicon pixel detector that typically provides four measurements per track, the first hit normally being in the insertable B-layer installed before \RunTwo~\cite{ATLAS-TDR-19,PIX-2018-001}, a silicon microstrip tracker, and a transition radiation tracker (TRT) that covers the region up to $\abseta = 2.0$. The TRT also provides electron identification information based on the fraction of hits (typically 30 in total) above a higher energy-deposit threshold corresponding to transition radiation.
 
Outside the thin superconducting solenoid, the calorimeter system covers the pseudorapidity range $|\eta| < 4.9$. Within the region $|\eta|< 3.2$, electromagnetic calorimetry is provided by barrel and endcap high-granularity lead/liquid-argon (LAr) calorimeters, with an additional thin LAr presampler covering $|\eta| < 1.8$ to correct for energy loss in material upstream of the calorimeters. Hadron calorimetry is provided by the steel/scintillator-tile calorimeter, segmented into three barrel structures within $|\eta| < 1.7$ and two copper/LAr hadron endcap calorimeters. The solid angle coverage is completed with forward copper/LAr and tungsten/LAr calorimeter modules optimised for electromagnetic and hadronic energy measurements respectively.
 
In the outer part of the ATLAS detector, the muon spectrometer (MS) covers the region $|\eta| < 2.7$ with three layers of precision tracking chambers. These consist of monitored drift tubes except in the innermost layer at $2.0 < |\eta| < 2.7$, where cathode-strip chambers are used to cope with the much higher rate of background particles.
The muon trigger system covers the range $|\eta| < 2.4$ with resistive-plate chambers in the barrel, and thin-gap chambers in the endcap regions. The MS is immersed in a magnetic field produced by three large superconducting air-core toroidal magnets with eight coils each. The field integral of the toroids ranges between \num{2.0} and \SI{6.0}{\tesla\metre} across most of the detector.
 
A two-level trigger system selects events that are of interest for the ATLAS physics programme~\cite{TRIG-2016-01}. The first-level trigger is implemented in hardware and reduces the event rate to below \SI{100}{\kHz}. A software-based trigger further reduces this to a recorded event rate of approximately \SI{1}{\kHz}.
 
An extensive software suite~\cite{ATL-SOFT-PUB-2021-001} is used in data simulation, in the reconstruction and analysis of real and simulated data, in detector operations, and in the trigger and data acquisition systems of the experiment.

\clearpage

\clearpage
\section{Dataset and simulated event samples}
\label{sec:monteCarloProduction}
 
This analysis uses data from \pp collisions occurring every \SI{25}{\ns} at $\sqrt{s}=\SI{13}{\TeV}$, collected during \RunTwo of the LHC from 2015 to 2018. Events selected for analysis are required to pass standard data-quality requirements~\cite{DAPR-2018-01} and they amount to a total integrated luminosity of \SI{139}{\ifb}. The uncertainty in the combined 2015--2018 integrated luminosity is \SI{1.7}{\%}~\cite{ATLAS-CONF-2019-021}, obtained using the LUCID-2 detector~\cite{LUCID2} for the primary luminosity measurements. Events were collected using two-lepton triggers that select pairs of electrons~\cite{TRIG-2018-05} or muons~\cite{TRIG-2018-01} or electron--muon combinations~\cite{TRIG-2018-05}. The transverse momentum (\pt) thresholds of the unprescaled two-lepton triggers were raised during \RunTwo because of the increasing luminosity of the colliding beams but were never higher than \SI{24}{\GeV} (\SI{24}{\GeV}) for the (sub-)leading electron or \SI{22}{\GeV} (\SI{14}{\GeV}) for the (sub-)leading muon.
 
Monte Carlo (MC) methods were used to model signal and background events. The \GEANT toolkit~\cite{Agostinelli:2002hh} was used to simulate the response of the ATLAS detector~\cite{SOFT-2010-01}. Simulated events were reconstructed with the same algorithms as those applied to data~\cite{ATL-SOFT-PUB-2021-001} and were corrected with calibration factors to better match the performance measured in data. The various tools and set-ups used to simulate the different processes are summarised in \Tab{\ref{tab:MC}}. To minimise the impact of the theoretical uncertainties on the final result, the normalisation of the MC samples that model the dominant backgrounds is considered a free parameter in the final likelihood fit, described in \Sect{\ref{sec:results}}.
 
Signal events were generated at leading order (\LO) with the LRSM package in \PYTHIA[8.212]~\cite{Sjostrand:2014zea}, which implements the \DCH scenario described in \Refn{\cite{huitu1997doubly}}. All samples contain a mixture of $\DCH_L$ and $\DCH_R$ particles. A set of tuned parameters called the A14 tune~\cite{ATL-PHYS-PUB-2014-021} and the \NNPDF[2.3lo] parton distribution function~(PDF) set~\cite{Ball:2012cx} were used. The value of $h_{\ell \ell^\prime}$ in \Eqn{(\ref{eq:decay-rate})} was set to $0.02$ for each leptonic decay mode to obtain a \DCH decay width that is negligible compared to the detector resolution. Consequently, the branching ratios are assumed to be equal for all possible leptonic final states. Only \DY production of the \DCH particles was simulated. The $\DCH_L$ pair production cross-sections at $\sqs=\SI{13}{\TeV}$ were calculated to leading-order and next-to-leading-order (\NLO) accuracies~\cite{Fuks:2019clu}. The resulting \NLO-to-\LO $K$-factor was then applied to the $\DCH_R$ \LO production cross-section, $\sigma_R^{\LO}$. The cross-sections for the Zee--Babu model were calculated using the same simulation set-up at \NLO accuracy~\cite{Ruiz:2022sct} and agree well with the $K \times \sigma_R^{\LO}$ value, as expected. Eleven samples were simulated with different masses of the \DCH particle decaying into light leptons, and an additional 11 samples with at least one $\tau$-lepton in the final state were produced. Signal mass points considered in the analysis range from \SI{400}{\GeV} to \SI{1300}{\GeV} in steps of \SI{100}{\GeV}.
 
\DY $\ell^+\ell^-$ production was modelled with the \SHERPA[2.2.1] generator~\cite{Bothmann:2019yzt} and the \NNPDF[3.0nnlo] set of PDFs~\cite{Ball:2014uwa}, using \NLO matrix elements for up to two partons, and \LO matrix elements for up to four partons, calculated with the Comix~\cite{Gleisberg:2008fv} and \OPENLOOPS~\cite{Buccioni:2019sur,Cascioli:2011va,Denner:2016kdg} libraries. They were matched with the \SHERPA parton shower~\cite{Schumann:2007mg} using the \MEPSatNLO prescription~\cite{Hoeche:2011fd,Hoeche:2012yf,Catani:2001cc,Hoeche:2009rj}. The parton shower used the set of tuned parameters developed by the \SHERPA authors, and the samples were normalised to a next-to-next-to-leading-order (\NNLO) prediction~\cite{Anastasiou:2003ds}. Similar methods were used to simulate \Wjets processes that contribute to the fake-lepton background, which is estimated in \Sect{\ref{sec:backgroundEstimation}}.
 
Samples of diboson ($VV$) and multiboson ($VVV$) final states were simulated with the \SHERPA~\cite{Bothmann:2019yzt} generator, using version 2.2.1 or 2.2.2, depending on the process. Where appropriate, off-shell effects and Higgs boson contributions were included. Fully leptonic final states and semileptonic final states, where one boson decays leptonically and the other hadronically, were generated using matrix elements at \NLO accuracy in QCD for up to one additional parton and at \LO accuracy for up to three additional parton emissions. Samples for the loop-induced processes, $gg \to VV$, were generated using \LO-accurate matrix elements for up to one additional parton emission for both the fully leptonic and semileptonic final states. The matrix element calculations were matched and merged with the \SHERPA parton showers, using Catani--Seymour dipole factorisation~\cite{Gleisberg:2008fv,Schumann:2007mg} with the \MEPSatNLO prescription~\cite{Hoeche:2011fd,Hoeche:2012yf,Catani:2001cc,Hoeche:2009rj}. The virtual QCD corrections were provided by the \OPENLOOPS library~\cite{Buccioni:2019sur,Cascioli:2011va,Denner:2016kdg}. The \NNPDF[3.0nnlo] set of PDFs was used~\cite{Ball:2014uwa}, along with the dedicated set of tuned parton-shower parameters developed by the \SHERPA authors.
 
\begin{table}[htbp]
\centering
\caption{Simulated signal and background event samples: the corresponding event generator, PDF set used for the matrix element, cross-section normalisation, set of tuned parameters and parton shower are shown for each sample. The generator cross-section is used where not specifically stated otherwise.}
\label{tab:MC}

\scriptsize
\resizebox{\textwidth}{!}{
\begin{tabular}{l l c c c c}
\toprule
Physics process &
Event generator &
PDF set         &
Cross-section   &
Parton shower   &
Parton shower   \\
& & & normalisation & & tune \\
\midrule
 
Signal & \multirow{2}{*}{\PYTHIA[8.212]\cite{Sjostrand:2014zea}} & \multirow{2}{*}{\NNPDF[2.3lo]~\cite{Ball:2012cx}} & \multirow{2}{*}{\NLO} & \PYTHIA[8.212] & \multirow{2}{*}{A14~\cite{ATL-PHYS-PUB-2014-021}} \\
\DCH & & & & \& \EVTGEN[1.6.0] & \\
 
\midrule
 
\DY & & & & & \\
$\Zboson/\gam^{*}\rightarrow \ee/\mumu/\tautau$ & \SHERPA[2.2.1]~\cite{Bothmann:2019yzt} & \NNPDF[3.0nnlo]~\cite{Ball:2014uwa} & \NNLO & \SHERPA & \SHERPA default \\
 
\midrule
Diboson & & & & & \\
$ZZ$, $WZ$, $WW$ & \SHERPA[2.2.1] \& 2.2.2 & \NNPDF[3.0nnlo] & \NLO & \SHERPA & \SHERPA default \\
 
Multiboson & & & & & \\
$WWW$, $WWZ$, $WZZ$, $ZZZ$ & \SHERPA[2.2.1] \& 2.2.2 & \NNPDF[3.0nnlo] & \NLO & \SHERPA & \SHERPA default \\
\midrule
 
Top-quark processes & & & & & \\
$\ttbar$ & \POWHEGBOX[v2]~\cite{Frixione:2007nw,Nason:2004rx,Frixione:2007vw,Alioli:2010xd} & \NNPDF[3.0nlo] & \NNLO + NNLL & \PYTHIA[8.230] & A14 \\
Single $t$ & \POWHEGBOX[v2]~\cite{Alioli:2010xd,Nason:2004rx,Frixione:2007vw} & \NNPDF[3.0nlo] & \NLO & \PYTHIA[8.230] & A14 \\
3$t$, 4$t$ & \MGNLO[2.2.2]~\cite{Alwall:2014hca} & \NNPDF[2.3lo] & \LO & \PYTHIA[8.186] & A14 \\
$\ttbar$ + $W/Z/H$ & \MGNLO[2.3.3] & \NNPDF[3.0nlo] & \NLO & \PYTHIA[8.210] & A14 \\
 
\bottomrule
\end{tabular}
}


\end{table}
 
The \NLO \POWHEGBOX[v2]~\cite{Frixione:2007nw,Nason:2004rx,Frixione:2007vw,Alioli:2010xd} generator was used with the \NNPDF[3.0nlo]~\cite{Ball:2014uwa} PDF set to model the production of \ttbar and single-top-quark $Wt$-channel events; the top-quark mass was set to \SI{172.5}{\GeV} and the \hdamp parameter\footnote{The \hdamp parameter is a resummation damping factor and one of the parameters that controls the matching of \POWHEG matrix elements to the parton shower and thus effectively regulates the high-\pt radiation against which the \ttbar system recoils.} was set to $1.5\,\mtop$~\cite{ATL-PHYS-PUB-2016-020}. For the $Wt$-channel single-top events, the diagram removal scheme~\cite{Frixione:2008yi} was used to remove interference and overlap with \ttbar production. The events were interfaced to \PYTHIA[8.230]~\cite{Sjostrand:2014zea} to model the parton shower, hadronisation, and underlying event, with parameters set according to the A14 tune~\cite{ATL-PHYS-PUB-2014-021} and using the \NNPDF[2.3lo] set of PDFs~\cite{Ball:2012cx}. The decays of bottom and charm hadrons were performed by \EVTGEN[1.6.0]~\cite{Lange:2001uf}. Additionally, top-quark spin correlations are preserved through the use of \textsc{MadSpin}~\cite{Artoisenet:2012st}. The predicted \ttbar production cross-section, calculated with \TOPpp[2.0]~\cite{Czakon:2011xx} to \NNLO in perturbative QCD and including soft-gluon resummation to next-to-next-to-leading-log order, is $830^{+20}_{-29}\,\mathrm{(scale)} \pm 35\,(\mathrm{PDF} + \alphas)\,\si{\pb}$.
 
The \MGNLO[2.3.3]~\cite{Alwall:2014hca} generator was used with the \NNPDF[3.0nlo]~\cite{Ball:2014uwa} PDF set to model the production of $\ttbar + W/Z/H$ events. The events were interfaced to \PYTHIA[8.210]~\cite{Sjostrand:2014zea} using the A14 tune~\cite{ATL-PHYS-PUB-2014-021} and the \NNPDF[2.3lo]~\cite{Ball:2014uwa} PDF set. The production of $3t$ and $4t$ events was modelled using the \MGNLO[2.2.2]~\cite{Alwall:2014hca} generator at \LO with the \NNPDF[2.3lo]~\cite{Ball:2014uwa} PDF set. The events were interfaced with \PYTHIA[8.186]~\cite{Sjostrand:2007gs} using the A14 tune~\cite{ATL-PHYS-PUB-2014-021} and the \NNPDF[2.3lo]~\cite{Ball:2014uwa} PDF set. The decays of bottom and charm hadrons were simulated using the \EVTGEN[1.2.0] program~\cite{Lange:2001uf}.
 
Top-quark processes (\ttbar, single-top-quark, $3t$, $4t$ and $\ttbar + W/Z/H$) and multiboson processes ($\Wboson\Wboson\Wboson$, $\Wboson\Wboson\Zboson$, $\Wboson\Zboson\Zboson$ and $\Zboson\Zboson\Zboson$) are grouped together and form the `Other' background category. Pile-up events from additional interactions in the same or neighbouring bunch crossings were simulated with \PYTHIA[8.186]~\cite{Sjostrand:2007gs} using the \NNPDF[2.3lo] PDFs and the A3 tune~\cite{ATL-PHYS-PUB-2016-017} and overlaid on the simulated hard-scatter events, which were then reweighted to match the pile-up distribution observed in data.
 
\FloatBarrier

\clearpage

\clearpage
\section{Event reconstruction and selection}
\label{sec:eventSelection}
 
To reduce non-collision backgrounds originating from beam-halo events and cosmic rays, events are required to have at least one reconstructed primary vertex with at least two associated tracks, each having \pt greater than \SI{500}{\MeV}. Among all the vertices in the event, the one with the highest sum of squared \pt of the associated tracks is identified as the primary vertex. Events that contain jets must also satisfy the quality criteria described in \Refn{\cite{ATLAS-CONF-2015-029}}.
 
\subsection{Event reconstruction}
\label{sec:ObjSelection}
 
This analysis classifies leptons into two exclusive categories called \textit{tight} and \textit{loose}, defined specifically for each lepton flavour as described in \Sect{\ref{sec:backgroundEstimation}}. Leptons selected in the \textit{tight} category are mostly prompt leptons that are estimated using Monte Carlo samples and originate from decays of \Zboson, \Wboson, $H$ bosons, or from prompt leptonic $\tau$ decays, while \textit{loose} leptons are mostly non-prompt and misidentified leptons used for the estimation of reducible background. Lepton definitions described in the following three paragraphs correspond to the \textit{tight} category and the corresponding selection is referred to as \textit{baseline event selection}.
 
Electron candidates are reconstructed by using electromagnetic calorimeter and ID information to match an isolated calorimeter energy deposit to a charged-particle track. They are required to stay within the fiducial volume of the inner detector, $\abseta < 2.47$, and to have transverse momenta $\pt > \SI{40}{\GeV}$. Moreover, electrons must pass at least the \texttt{LHTight} identification level, based on a multivariate likelihood discriminant~\cite{EGAM-2018-01,PERF-2017-01}. The discriminant relies on track and calorimeter energy cluster information. Electron candidates within the transition region between the barrel and endcap electromagnetic calorimeters ($1.37 < \abseta < 1.52$) are vetoed due to limitations in their reconstruction quality. The track associated with the electron candidate must have a transverse impact parameter \dzero that, evaluated at the point of closest approach of the track to the beam axis in the transverse plane, satisfies $|\dzero|/\sigma(\dzero) < 5$, where $\sigma(\dzero)$ is the uncertainty on \dzero. In addition, the longitudinal distance \zzero from the primary vertex to the point where \dzero is measured must satisfy $|\zzero\sin(\theta)| < \SI{0.5}{mm}$, where $\theta$ is the polar angle of the track. The combined identification and reconstruction efficiency for \texttt{LHTight} electrons varies from \SI{58}{\%} to \SI{88}{\%} over the transverse energy range from \SI{4.5}{\GeV} to \SI{100}{\GeV}. Electron candidates must also satisfy the \texttt{FCLoose} isolation requirement, which uses calorimeter-based and track-based isolation criteria that reach an efficiency of about \SI{99}{\%}. Electron candidates are discarded if their angular distance from a jet satisfies $0.2 < \DeltaR < 0.4$. If $\DeltaR(e,j) < 0.2$, the jet is discarded. Both the reconstruction and isolation performance are evaluated in \Zee decay measurements described in \Refn{\cite{EGAM-2018-01}}.
 
A dedicated machine-learning algorithm (BDT) is applied to reject electrons with incorrectly identified charge~\cite{EGAM-2018-01}. A selection requirement on the BDT output is chosen to achieve a rejection factor that is between $7$ and $10$ for electrons having a wrong charge assignment while selecting properly measured electrons with an efficiency of \SI{97}{\%}.
 
Muon candidates are reconstructed by matching a track from the muon spectrometer with an inner-detector track. The candidates must have $\pt > \SI{40}{\GeV}$ and $\abseta < 2.5$, and satisfy the impact parameter requirements $|\dzero|/\sigma(\dzero) < 3$ and $|\zzero\sin(\theta)| < \SI{0.5}{mm}$. If a muon candidate has a \pt higher than \SI{300}{\GeV}, it must satisfy a special \texttt{HighPt} muon quality requirement. If its \pt is less than \SI{300}{\GeV}, the \texttt{Medium} quality requirement is used. The criteria associated with these selections are described in \Refn{\cite{MUON-2018-03}}. The muon candidates must also meet the track-based \texttt{FixedCutTightTrackOnly} isolation requirement. The full set of selections results in a reconstruction and identification efficiency of over \SI{95}{\%} in the entire phase space, as measured in \Zmm events~\cite{PERF-2015-10}. If a muon and a jet featuring more than three tracks lie $\Delta R < 0.4$ apart and the muon's transverse momentum is less than half of the jet's \pt, the muon is discarded. Furthermore, a muon candidate that deposits a sufficiently large fraction of its energy in the calorimeter is rejected if it also shares a reconstructed ID track with an electron.
 
Jets are reconstructed using particle-flow objects that combine tracking and calorimetric information~\cite{PERF-2015-09}. The anti-$k_{t}$ algorithm~\cite{Cacciari:2008}, implemented in \textsc{FastJet}~\cite{Cacciari:2011ma}, is applied with a radius parameter $R = 0.4$. The measurements of jet \pt are calibrated to the particle level~\cite{JETM-2018-05}. To help reject additional jets produced by \pileup, a `jet vertex tagger' (JVT) algorithm~\cite{ATLAS-CONF-2014-018} is used for jets with $\pt < \SI{60}{GeV}$ and $\abseta < 2.4$. It employs a multivariate technique that relies on jet energy and tracking variables to determine the likelihood that the jet originates from the primary vertex. The \textit{medium} JVT working point is used, which has an average efficiency of \SI{92}{\%}.
 
Jets considered in this analysis are required to have $\pt > \SI{20}{GeV}$ and $\abseta < 2.5$. Jets within $\Delta R = 0.2$ of an electron are discarded. Jets within $\Delta R = 0.2$ of a muon and featuring fewer than three tracks or having $\pt(\mu) / \pt(\mathrm{jet}) > 0.5$ are also discarded. Events containing jets identified as originating from $b$-quarks are vetoed. They are identified with a multivariate algorithm~\cite{JETM-2018-05} which uses information about the impact parameters of inner-detector tracks matched to the jet, the presence of displaced secondary vertices, and the reconstructed flight paths of $b$- and $c$-hadrons inside the jet~\cite{FTAG-2018-01,ATLAS-CONF-2018-006,ATL-PHYS-PUB-2017-013}. The algorithm is used at a working point that provides a $b$-tagging efficiency of \SI{77}{\%}, as determined in a sample of simulated \ttbar events. The working point provides rejection factors of approximately $134$, $6$, and $22$ for light-quark and gluon jets, $c$-jets, and hadronically decaying $\tau$-leptons, respectively.
 
The missing transverse momentum is defined as the negative vector sum of the transverse momenta of fully reconstructed particles and its magnitude is denoted by \MET. If an event has tracks that are not assigned to any of the physics objects but are matched to the primary vertex, an additional `soft term' is added. In a high \pileup environment, this term improves the reconstruction of \MET~\cite{ATLAS-CONF-2018-023}.
 
Correction factors are applied to the selected electrons, muons, and jets. For electrons and muons, these factors account for identification and selection efficiency differences between data and simulation~\cite{EGAM-2018-01,MUON-2018-03}, while for jets, they correct the jet energy scale and jet energy resolution~\cite{JETM-2018-05}.

\subsection{Event selection}
 
The baseline event selection requires at least two light leptons to be identified as \textit{tight}. For this search, three distinct types of regions are defined: \textit{control regions} (CR), \textit{validation regions} (VR), and \textit{signal regions} (SR). The normalisation factors of the dominant backgrounds are treated as free parameters in the likelihood fit and are constrained by the control regions. Validation regions are used to cross-check the fit results but are not included in the fit itself. Both the control regions and validation regions are intended to be close to the kinematic region of the expected signal but must be constructed in a way to allow only negligible signal contamination. The CRs and VRs are defined using selections that are orthogonal to the SRs. The analysis is sensitive to final states with three different lepton multiplicities. The regions are optimised to search for one same-charge lepton pair in the case of regions requiring two or three leptons, while four-lepton events are required to feature two same-charge pairs with zero total charge. The main variable used to distinguish between control, validation and signal regions is the invariant mass of the two same-charge leptons with the highest \pt in the event, \mLL, where $\ell, \ell^\prime=e,~\mu$. This variable is also used in the final fit to data in the two- and three-lepton regions, while a single-bin event yield is used in the four-lepton regions. The selection criteria for each region are summarised in Table~\ref{tab:regions}.
 
\begin{table}[t]
\centering
\caption{Summary of regions defined in the analysis. The table is split into four horizontal blocks: the upper block indicates the final states for each region, the second block lists the lepton multiplicity of the region, the third block indicates the mass range of the corresponding final state, and the lower block indicates the event selection criteria for the region. The application of a selection requirement is indicated by a check-mark (\ding{51}), or by \textit{inverted} when it is inverted. The three- and four-lepton regions include all light-lepton flavour combinations. No selection is applied when a dash is present in the corresponding cell. A selection on the average invariant mass of the two same-charge lepton pairs, $\overline{m} \equiv (m_{\ell^+\ell^{\prime +}}+m_{\ell^-\ell^{\prime -}})/2$, is used to increase the signal significance in the four-lepton signal region.}
\label{tab:regions}

\resizebox{\textwidth}{!}{
\begin{tabular}{ c V{4} c | c c | c | c V{4} c | c | c V{4} c c | c | c }
& \multicolumn{5}{c V{4}}{Control regions} & \multicolumn{3}{c V{4}}{Signal regions} & \multicolumn{4}{c}{Validation regions} \\
\toprule
& \texttt{DYCR} & \multicolumn{2}{c|}{\texttt{DBCR2L}} & \texttt{DBCR3L} & \texttt{CR4L} & \texttt{SR2L} & \texttt{SR3L} & \texttt{SR4L} & \multicolumn{2}{c|}{\texttt{VR2L}} & \texttt{VR3L} & \texttt{VR4L} \\
\toprule
\multirow{3}{*}{Channel} & $e^+e^-$ & $e^\pm e^\pm$ & & $\ell^\pm \ell^\pm \ell^\mp$ & $\ell^+ \ell^+ \ell^- \ell^-$ & $e^\pm e^\pm$ & $\ell^\pm \ell^\pm \ell^\mp$ & $\ell^+ \ell^+ \ell^- \ell^-$ & $e^\pm e^\pm$ & & $\ell^\pm \ell^\pm \ell^\mp$ & $\ell^+ \ell^+ \ell^- \ell^-$ \\
& & & $e^\pm \mu^\pm$ & & & $e^\pm \mu^\pm$ & & & & $e^\pm \mu^\pm$ & & \\
& & & $\mu^\pm \mu^\pm$ & & & $\mu^\pm \mu^\pm$ & & & & $\mu^\pm \mu^\pm$ & & \\
\midrule
Number of leptons & $2$ & \multicolumn{2}{c|}{$2$} & $3$ & $4$ & $2$ & $3$ & $4$ & \multicolumn{2}{c|}{$2$} & $3$ & $4$ \\
\midrule
$m(\ell^{\pm}, \ell^{\prime \mp})_\mathrm{lead}$ [\si{\GeV}] & $\ge 300$ & \multicolumn{2}{c|}{-} & - & - & - & - & - & \multicolumn{2}{c|}{-} & - & - \\
 
\mLL[\si{\GeV}] & - & \multicolumn{2}{c|}{$[200,~300)$} & $\ge 300$ & $[100,~200)$ &  $\ge 300$ & $\ge 300$ & $\ge 300$ & \multicolumn{2}{c|}{$\ge 300$} & $[100,~300)$ & $[200,~300)$ \\
 
\midrule
 
\PtPairLep[\si{\GeV}] & - & \multicolumn{2}{c|}{-} & - & - & $\ge 300$ & $\ge 300$ & - & \multicolumn{2}{c|}{$[200,~300)$} & - & - \\
 
\DRLep & - & \multicolumn{2}{c|}{-} & - & - & $< 3.5$ & - & - & \multicolumn{2}{c|}{$< 3.5$} & - & - \\
 
$\overline{m}$ [\si{\GeV}] & - & \multicolumn{2}{c|}{-} & - & - & - & - & $\ge 300$ & \multicolumn{2}{c|}{-} & - & - \\
 
\MET[\si{\GeV}]  & - & $> 30$ & - & - & - & - & - & - & $> 30$ & - & - & - \\
 
$|\eta(\ell,\ell^\prime)|$  & - & $< 3.0$ & - & - & - & - & - & - & $< 3.0$ & - & - & - \\
 
\Zboson-veto & - & \multicolumn{2}{c|}{-} & inverted & - & - & \ding{51} & \ding{51} & \multicolumn{2}{c|}{-} & \ding{51} & - \\
 
\bottomrule
\end{tabular}
}


\end{table}
 
Events with at least one $b$-tagged jet are vetoed in all regions to suppress background events arising from top-quark decays. In regions with more than two leptons, the so-called \Zboson-veto condition is required, i.e.\ events containing same-flavour lepton pairs having invariant masses within \SI{20}{\GeV} of the \Zboson-boson mass ($\SI{71.2}{\GeV} < m(\ell, \ell) < \SI{111.2}{\GeV}$) are rejected in order to suppress events featuring a \Zboson boson in the final state. The \Zboson-veto is not applied in four-lepton control and validation regions -- this increases the available number of simulated diboson events.
 
\subsubsection{Signal regions}
The signal regions, independently of the lepton multiplicity and flavour combination, require invariant masses of the same-charge leading lepton pair to be above \SI{300}{\GeV}. To maximise the sensitivity, additional requirements are imposed on same-charge lepton pairs, regardless of the flavour. These exploit both the boosted decay topology of the \DCH resonance and the high energy of the decay products. The same-charge lepton angular separation in the two-lepton signal region (SR2L) is required to be $\Delta R(\ell^{\pm},\ell^{\prime \pm}) < 3.5$. Furthermore, the vector sum of the two leading leptons' transverse momenta, \PtPairLep, must exceed \SI{300}{\GeV} in both the two-lepton and three-lepton (SR3L) signal regions. In the four-lepton signal region (SR4L), the signal significance is increased by requiring the average invariant mass of the two same-charge lepton pairs to satisfy $\overline{m} = (m_{\ell^+\ell^{\prime +}}+m_{\ell^-\ell^{\prime -}})/2 > \SI{300}{\GeV}$.
 
The efficiency for signal events to pass the signal region selections depends on the signal mass and on the flavour combination of the leptons in the final states. For the benchmark \DCH masses, the efficiencies range between \SI{10}{\%} and \SI{20}{\%} in the two-lepton signal regions, while the typical efficiencies in the three- and four-lepton signal regions vary from \SI{30}{\%} to \SI{40}{\%}. Generally, the efficiencies increase with the \DCH mass. Focusing on two-lepton regions, the efficiencies are highest for the $e \mu$ channel, lower for the $ee$ channel and lowest for the $\mu \mu$ channel, which is an effect of the two-lepton triggers used in the search.
 
The following control and validation regions either span a lower \mLL interval or require other orthogonal selections if the same \mLL window as in the signal regions is used.
 
\subsubsection{Control regions}
Firstly, the two-lepton diboson control region (DBCR2L) is used to constrain the dominant diboson contribution in the two-lepton final states and spans a $\mLL \in [200,~300)~\si{\GeV}$ range. A significant \DY event yield was observed in the $ee$ channel, so additional selections on $\MET$ and the pseudorapidity of the same-charge lepton pair, $|\eta(\ell^\pm, \ell^{\prime \pm})|$, are required to specifically suppress the \DY background. The \DY background is further constrained by the opposite-charge \DY control region (DYCR), where exactly two electrons with opposite charges are required. Since it is designed to target only opposite-charge electron pairs, the invariant mass of the opposite-charge electron pair is required to be $m(e^{\pm},e^{\mp})_{\mathrm{lead}} > \SI{300}{\GeV}$ and is also used as the fit variable. Then, the three-lepton diboson control region (DBCR3L) is used to constrain the diboson background yield, independently of the flavour combination. Since the \mLL requirement is the same as it is in the corresponding three-lepton signal region, at least one \Zboson-boson candidate is required in order to achieve orthogonality. Finally, the four-lepton control region (CR4L) is used to constrain the yield of the dominant diboson background in four-lepton regions, where \mLL is restricted to be between \SI{100}{\GeV} and \SI{200}{\GeV}. The signal contamination in all control regions is negligible.
 
\subsubsection{Validation regions}
The two-lepton validation region (VR2L) is used to validate the data-driven fake-lepton background estimation and to assess the diboson modelling in the two-lepton channel. The \mLL selection is the same as in the SR2L, thus requiring an inverted \PtPairLep cut with respect to the corresponding signal region. Similarly to the DBCR2L, additional requirements on $\MET$ and $|\eta(\ell^\pm, \ell^{\prime \pm})|$ are imposed only in the $ee$ channel. The three-lepton validation region (VR3L) is used to validate the diboson background and fake-lepton background with three reconstructed leptons in the final states. The \mLL value is required to be within the interval of $[100,~300)~\si{\GeV}$. Additionally, a \Zboson-veto condition is applied. The four-lepton validation region (VR4L) is used to validate the diboson modelling in the four-lepton region and is defined by the $\SI{200}{\GeV} < \mLL < \SI{300}{\GeV}$ requirement. The signal yield does not exceed a few per cent of the background yield in any validation region.
 
\FloatBarrier

\clearpage

\clearpage
\section{Background composition and estimation}
\label{sec:backgroundEstimation}
 
In this section, background estimation techniques are described. Different lepton multiplicity and flavour channels in the analysis have different background compositions and thus require different treatments. Backgrounds can be categorised into irreducible and reducible types, which can be identified by their source. The former category is derived from MC simulation and the latter with data-driven methods.
 
Irreducible background sources are SM processes producing the same prompt final-state lepton pairs as the signal, with the dominant contributions in this analysis coming from diboson production. Irreducible prompt SM backgrounds in all regions are estimated using the simulated samples listed in \Sect{\ref{sec:monteCarloProduction}}.
 
Reducible background sources are processes where reconstructed leptons originate from misreconstructed objects such as jets, or from light- or heavy-quark decays or, in the electron case, from photon conversions. These types of events thus contain at least one non-prompt lepton. Events containing leptons whose charges were incorrectly assigned also enter the reducible background category. An example of this type of background is the \DY background, where the contribution is normalised to the DYCR, then reweighted for the charge misidentification probability and finally used to predict yields in the same-charge selections.
 
To avoid overlap between the irreducible backgrounds estimated using MC simulation and the data-driven, reducible backgrounds, events from the MC background samples are considered only if reconstructed leptons can be matched to their prompt generator-level counterparts. Electrons are the only significant source of charge misidentification.
 
\subsection{Electron charge misidentification}
 
Electron charge misidentification is caused mainly by bremsstrahlung emission from the electrons as they propagate through the detector material. The emitted photon can either convert to an electron--positron pair or traverse the inner detector without creating any track. In the first case, the calorimeter energy cluster corresponding to the initial electron can be matched to the wrong-charge track. In the case of photon emission without subsequent pair production, the electron track usually has only very few hits in the silicon pixel layers, while other hits are from unknown sources. This can lead to the wrong charge assignment from the track curvature, but a correct estimate of the electron energy, since its estimation relies mostly on the calorimeter.
 
The reducible background due to charge misidentification is estimated in same-charge analysis channels that contain electrons. The modelling of charge misidentification in the \GEANT simulations deviates from data because of the complex processes involved, which depend sensitively on the arrangement of the material in the detector. Consequently, charge reconstruction correction factors are derived by comparing the charge misidentification probability measured in \Zee data with the one in simulation. The charge misidentification probability is extracted by performing a likelihood fit to a dedicated \Zee data sample, as described in \Refn{\cite{EGAM-2018-01}}. These scale factors are then applied to the simulated background events to compensate for the differences.
 
\subsection{Fake/non-prompt lepton background}
 
The \textit{non-prompt} lepton background refers to a reducible background that originates from secondary decays of light- or heavy-flavour mesons into light leptons that are usually surrounded by jets. Another source of reducible background (\textit{fakes}) arises from other physics objects misidentified as electrons or muons. Collectively, such events are referred to as the \textit{fake/non-prompt} (FNP) background. The $b$-jet veto significantly reduces the number of FNP leptons from heavy-flavour decays. Most of the FNP leptons still passing the analysis selection originate from in-flight decays of mesons inside jets, jets misreconstructed as electrons, and conversions of initial- and final-state radiation photons. MC samples are not used to estimate these background sources because the simulation of jet production and hadronisation has large intrinsic uncertainties. It is estimated with a data-driven approach, the so-called \textit{fake factor} method described in \Refn{\cite{EXOT-2012-25}}. This method relies on determining the probability, or fake factor ($F_\ell$), for a FNP lepton to be identified as a prompt lepton, corresponding to the default lepton selection. Electrons that pass the \texttt{LHLoose} identification requirements~\cite{EGAM-2018-01,PERF-2017-01} but fail either the \texttt{LHTight} identification or the isolation requirements are referred to as \textit{loose}. Muons, on the other hand, must pass the \texttt{LHMedium} identification requirements~\cite{MUON-2018-03} but fail isolation to enter the \textit{loose} category. Electron and muon fake factors, $F_\ell$, are then defined as the ratio of the number of \textit{tight} leptons to the number of \textit{loose} leptons and are parameterised as functions of \pt, $\MET$, and $\eta$. The FNP lepton background (containing at least one fake lepton) is estimated in the SR by applying $F_\ell$ as a normalisation correction to relevant distributions in a region which has the same selection criteria as the SR, except that at least one of the two leptons must pass the \textit{loose} selection but fail the \textit{tight} one. The fake factor is measured in dedicated fake-enriched regions, reported in \Tab{\ref{tab:FakeEnrichedRegions}}, where events must pass prescaled support single-lepton triggers without isolation criteria, contain no $b$-jets and have exactly one reconstructed lepton that satisfies either the \textit{tight} selection criteria or a relaxed \textit{loose} selection. No additional requirements are used in the electron channel. Exactly one jet with a $\pt > \SI{35}{\GeV}$ and one reconstructed muon have to be back-to-back in the muon fake-enriched region. This is achieved by requiring $\Delta\phi(\mu,\mathrm{jet}) > 2.7$. In addition, $\Wboson \rightarrow \mu \nu$ events are rejected by requiring $\MET < \SI{40}{\GeV}$. Most of the selected events are dijet events, while the majority of prompt leptons originate from the \Wjets process.
 
\begin{table}[htbp]
\centering
\caption{Selection criteria defining the FNP lepton regions used to measure the ratio of the numbers of \textit{tight} and \textit{loose} leptons, the so-called fake factor, for the electron and muon channels. The fake factors are parameterised with a variable binning in \pt, which ensures that the statistical uncertainty stays below a fixed threshold.}
\label{tab:FakeEnrichedRegions}

\begin{tabular}{c|c}
 
\toprule
 
Muon channel & Electron channel \\
 
\midrule
 
\multicolumn{2}{c}{\pt variable binning} \\
\multicolumn{2}{c}{\bjet veto} \\
Single-muon trigger & Single-electron trigger \\
Exactly one muon & Exactly one electron \\
 
Five $\eta$ bins, two \MET bins & Four $\eta$ bins, two \MET bins \\
$\pt(\mathrm{jet}) > \SI{35}{\GeV}$ &  \\
$\Delta\phi(\mu,\mathrm{jet}) > 2.7$ & \\
\MET < \SI{40}{\GeV} & \\
 
\bottomrule
\end{tabular}


\end{table}
 
The fake-factor method relies on the assumption that no prompt leptons appear in the fake-enriched samples, which is not fully correct given the imposed selections. Therefore, the number of residual prompt leptons in the fake-enriched regions is estimated using the Monte Carlo samples described in \Sect{\ref{sec:monteCarloProduction}} and subtracted from the numbers of \textit{tight} and \textit{loose} leptons used to measure the fake factors.
 
\begin{figure}[htbp]
\centering
\subfloat[]{
\includegraphics[width=0.45\textwidth]{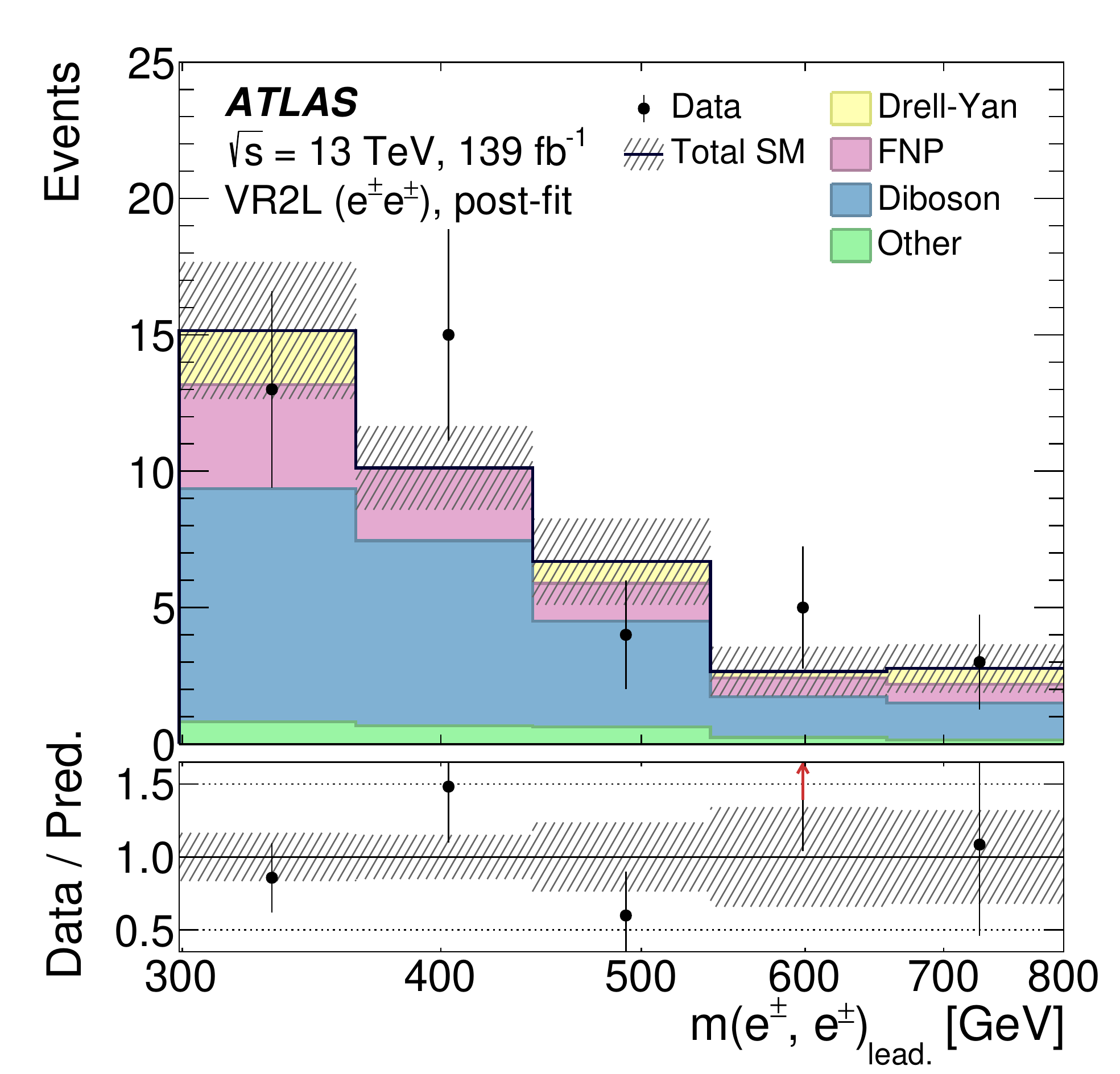}
\label{fig:histo_VR2Lee}
}
\subfloat[]{
\includegraphics[width=0.45\textwidth]{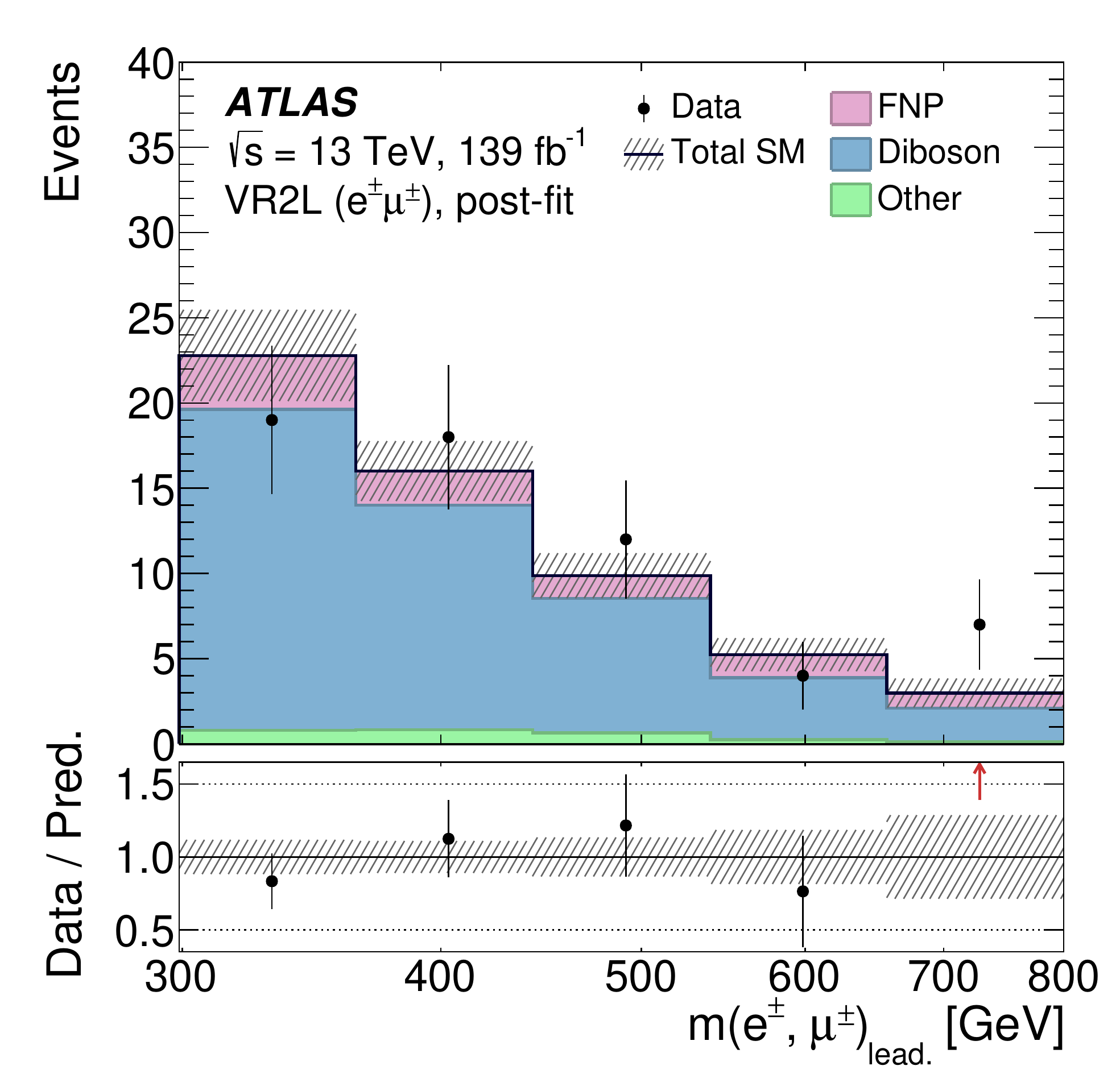}
\label{fig:histo_VR2Lem}
}
\\
\subfloat[]{
\includegraphics[width= 0.45\textwidth]{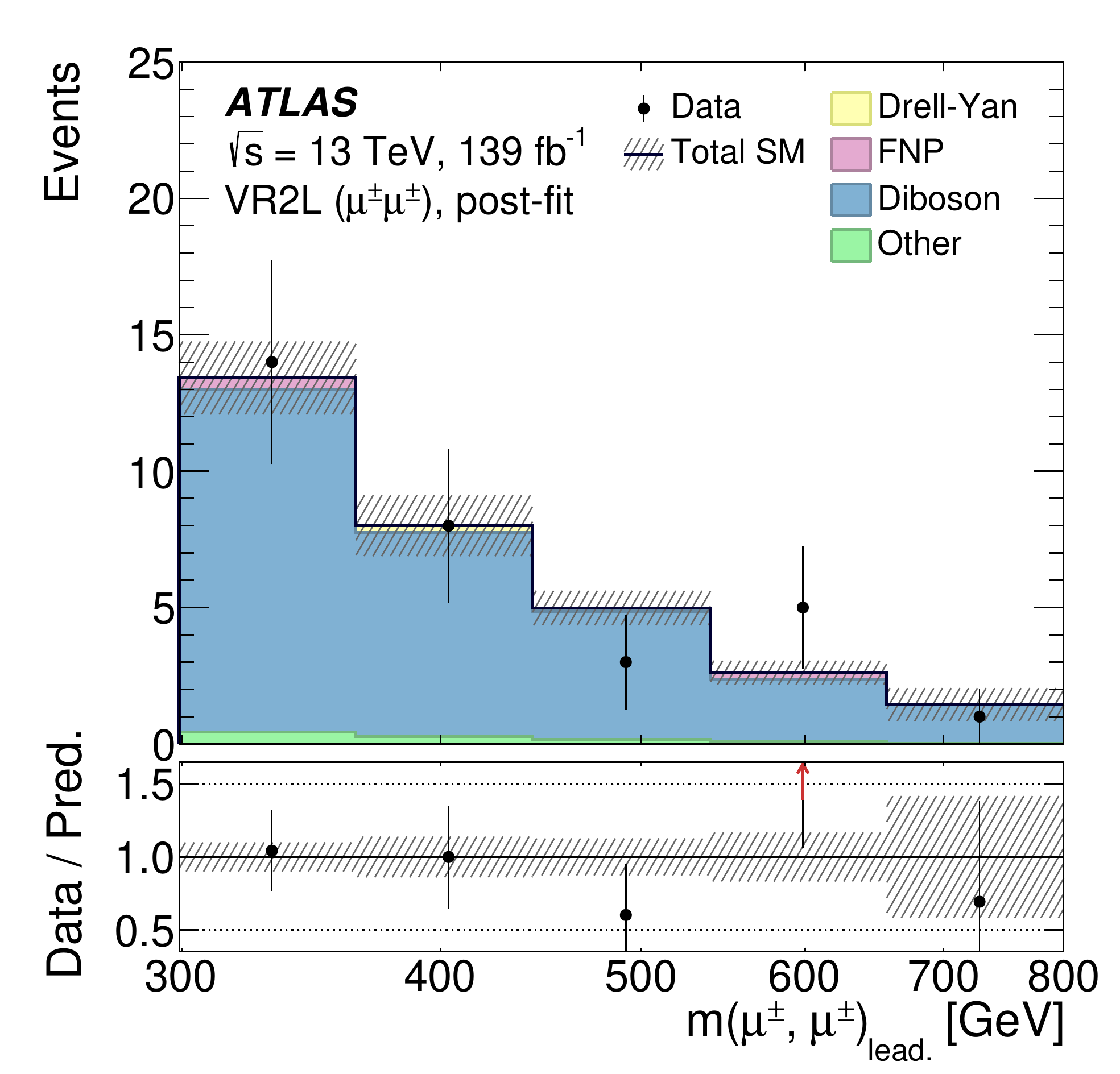}
\label{fig:histo_VR2Lmm}
}
 
\caption{Distributions of two-lepton mass \mLL for data and SM background predictions in validation regions:
\protect\subref{fig:histo_VR2Lee} the electron--electron,
\protect\subref{fig:histo_VR2Lem} the electron--muon, and
\protect\subref{fig:histo_VR2Lmm} the muon--muon two-lepton
validation regions. Backgrounds from top-quark and multiboson processes are merged, forming the `Other' category. The hatched bands include all systematic uncertainties after a background-only fit to data (post-fit), with the correlations between various sources taken into account. The last bin also includes any overflow entries. The error bars show statistical uncertainties. The lower panels show the ratio of the observed data to the estimated SM background. The red arrows indicate points that are outside the vertical range of the figure.}
\label{fig:representativeVRs}
\end{figure}

\begin{figure}[htbp]
\centering
\subfloat[]{
\includegraphics[width= 0.45\textwidth]{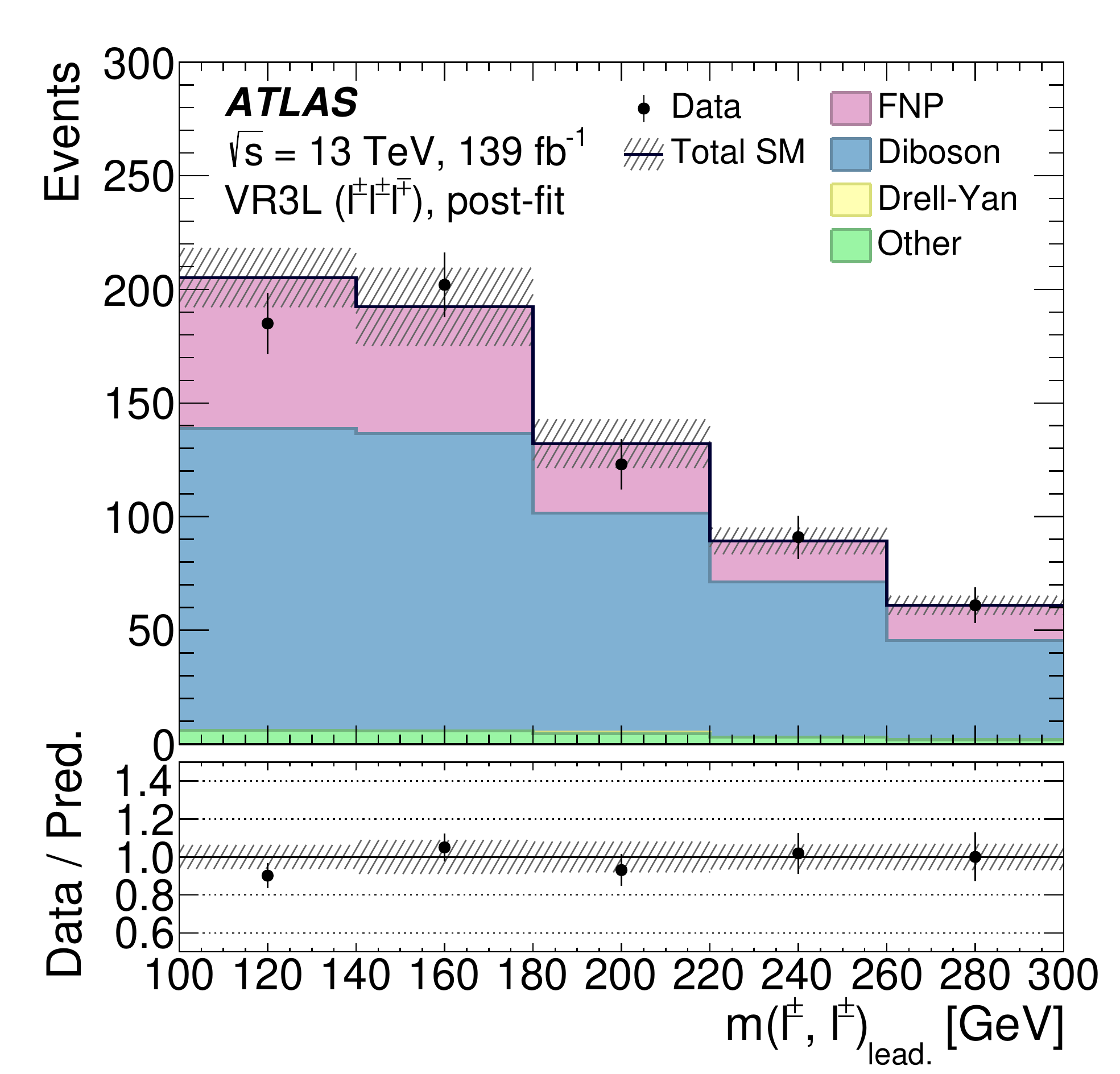}
\label{fig:histo_VR3L}
}
\subfloat[]{
\includegraphics[width= 0.45\textwidth]{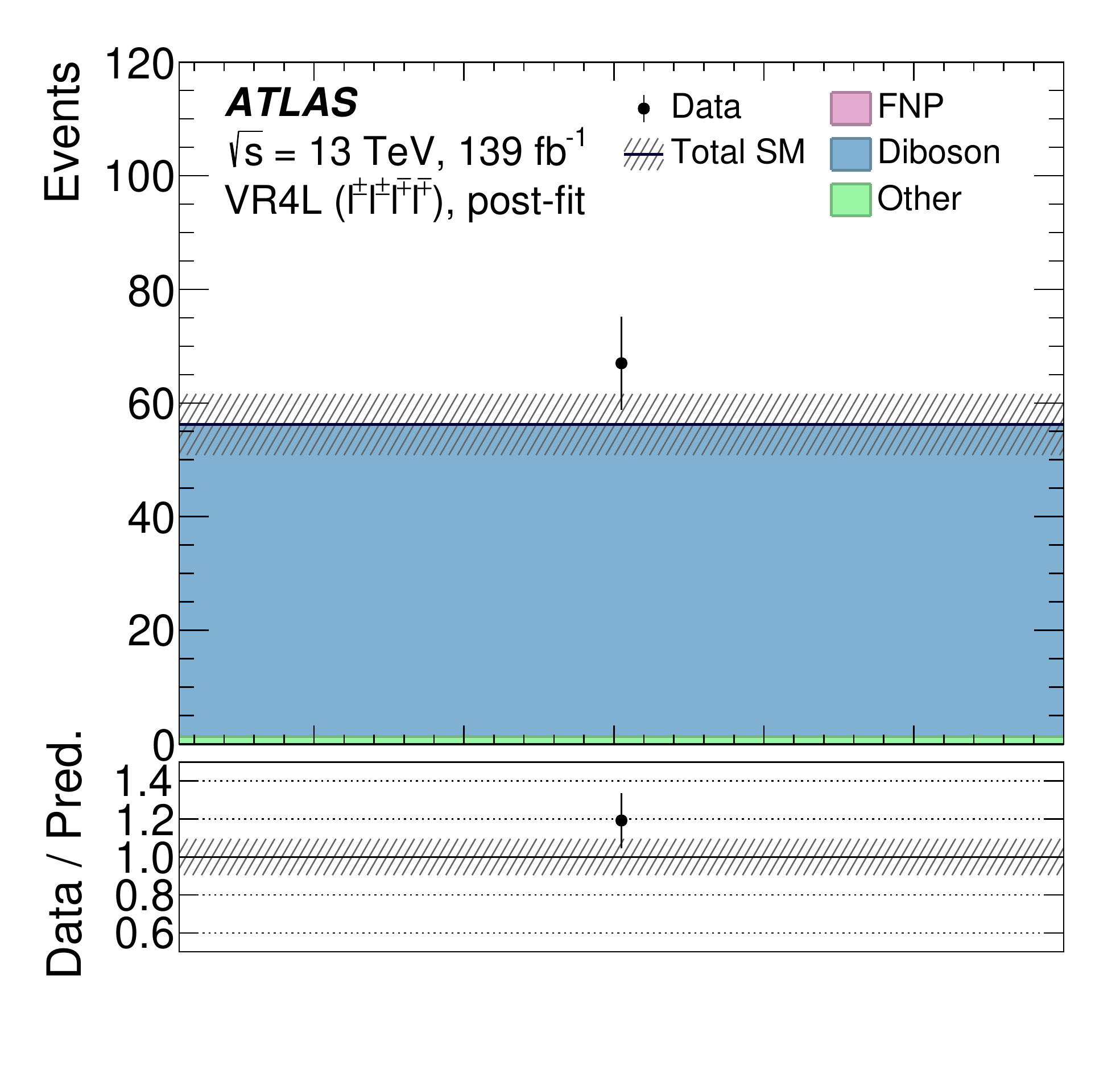}
\label{fig:histo_VR4L}
}
 
\caption{\protect\subref{fig:histo_VR3L} Distribution of two-lepton mass \mLL, where $\ell, \ell^\prime=e,~\mu$ for data and SM background predictions in three-lepton validation regions and \protect\subref{fig:histo_VR4L} event yield in the four-lepton validation region. Backgrounds from top-quark and multiboson processes are merged, forming the `Other' category. The hatched bands include all systematic uncertainties after a background-only fit to data (post-fit), with the correlations between various sources taken into account. The last bin also includes any overflow entries. The error bars show statistical uncertainties. The lower panels show the ratio of the observed data to the estimated SM background.}
\label{fig:representativeVRs2}
\end{figure}

Dedicated two-, three-, and four-lepton validation regions, defined in \Tab{\ref{tab:regions}}, are used to validate both the data-driven FNP lepton estimation and the modelling of the dominant diboson background in regions as similar to the signal regions as possible. \Fig{\ref{fig:representativeVRs}} and \Fig{\ref{fig:representativeVRs2}} present the \mLL distributions ($\ell, \ell^\prime=e,~\mu$) in the validation regions, which are sensitive to different background sources. A simultaneous background-only fit to data in all CRs was performed with diboson and \DY background normalisations considered as free parameters and with nuisance parameters corresponding to systematics listed in \Sect{\ref{sec:systematics}} included. Good background modelling is observed in all these regions.
 
\FloatBarrier
 
\section{Systematic uncertainties}
\label{sec:systematics}
 
Several sources of systematic uncertainty are accounted for in the analysis. These correspond to experimental and theoretical sources affecting both the background and signal predictions. All considered sources of systematic uncertainty affect the total event yield, and all, except the luminosity and cross-section uncertainties, also affect the distributions of the variables used in the fit (\Sect{\ref{sec:results}}).
 
The cross-sections used to normalise the simulated samples are varied to account for the scale and PDF uncertainties in the cross-section calculation. The variation is \SI{6}{\%} for diboson production~\cite{ATL-PHYS-PUB-2017-005}, \SI{13}{\%} for $\ttbar\,\Wboson$ production, \SI{12}{\%} for $\ttbar\,\Zboson$ production, and \SI{8}{\%} for $\ttbar\,H$ production~\cite{YR4}. Since the yield of the diboson background is derived from the likelihood fit to the data, these systematic variations contribute by changing the shapes and the relative normalization of the background predictions used in the likelihood fit in the CRs and SRs. The theoretical uncertainty in the \DY background is estimated by PDF eigenvector variations of the nominal PDF set, and variations of the PDF scale, \alphas, electroweak corrections, and photon-induced corrections. For \SHERPA-simulated processes (\DY, diboson and multiboson processes as listed in \Tab{\ref{tab:MC}}), uncertainties from missing higher orders were evaluated~\cite{Bothmann:2016nao} using seven variations of the QCD factorisation and renormalisation scales in the matrix elements by factors of $0.5$ and $2$, avoiding variations in opposite directions. The effect of the PDF choice is considered using the PDF4LHC prescription~\cite{Butterworth:2015oua}. Uncertainties in the nominal PDF set were evaluated using 100 replica variations. Additionally, the results were cross-checked using the central values of the \CT[14nnlo]~\cite{Dulat:2015mca} and \MMHT[nnlo]~\cite{Harland-Lang:2014zoa} PDF sets. The effect of the uncertainty in the strong coupling constant, \alphas, was assessed by variations of $\pm 0.001$. For \ttbar and single-top-quark samples, the uncertainties in the cross-section due to the PDF and \alphas were calculated using the \PDFforLHC[15] prescription~\cite{Butterworth:2015oua} with the \MSTW[nnlo]~\cite{Martin:2009iq,Martin:2009bu}, \CT[10nnlo]~\cite{Lai:2010vv,Gao:2013xoa} and \NNPDF[2.3lo]~\cite{Ball:2012cx} PDF sets in the five-flavour scheme, and were added in quadrature to the effect of the factorisation scale uncertainty. For \ttV production and processes producing three or more top quarks, the uncertainty due to initial-state radiation (ISR) was estimated by comparing the nominal event sample with two samples where the up/down variations of the A14 tune were employed. The theoretical uncertainty of the \NLO cross-section for $\pp \rightarrow H^{++}H^{--}$ is reported to range from a few per cent at low \DCH masses to approximately \SI{25}{\%}~\cite{Fuks:2019clu,Ruiz:2022sct} for the highest signal mass points studied in the analysis. It is not included in a fit as a nuisance parameter but is drawn as an uncertainty band around the theoretical curves in the exclusion limit plots in \Figs{\ref{fig:limit_xsec_DCH_full}}{{\ref{fig:limit_channels_DCH_full}}}. The uncertainty on signal cross-section includes the renormalisation and factorisation scale dependence and the uncertainty in the parton densities. The theoretical uncertainty in the $pp \rightarrow H^{++}H^{--}$ simulation is assessed by varying the A14 parameter set in \PYTHIA[8.186] and choosing CTEQ6L1 and CT09MC1~\cite{Lai:2009ne} as alternative PDFs. The impact on signal acceptance is found to be negligible. These theoretical uncertainties are considered fully correlated across the various analysis regions.
 
A significant contribution to the total uncertainty arises from the statistical uncertainty of the MC samples in all analysis regions. Analysis regions have very restrictive selections, and only a small fraction of the initially generated MC events remains after applying all requirements. The statistical uncertainty varies from \SI{5}{\%} to \SI{40}{\%} depending on the signal region.
 
Experimental systematic uncertainties due to differences between data and Monte Carlo for lepton reconstruction, identification, isolation, and trigger efficiencies are estimated by varying the corresponding scale factors. The impact of scale-factor systematic variations on the background yield estimation is at most \SI{3}{\%} and is less significant than the other systematic uncertainties and MC statistical uncertainties. The same is true for the lepton energy or momentum calibration.
 
Experimental systematic uncertainties due to different reconstruction and $b$-tagging efficiencies for reconstructed jets in data and simulation are taken into account. Uncertainties in the absolute jet energy scale and resolution, measured in multi-jet, \Zjets, and \gammajet events, are considered and are estimated to be less than \SI{2}{\%} across the range of $\pt(\mathrm{jet})$ of interest~\cite{JETM-2018-05,ATLAS-CONF-2018-023}. $b$-tagging efficiencies are measured in two-lepton \ttbar events and are estimated to range from \SI{8}{\%} at low momentum to \SI{1}{\%} at high momentum~\cite{FTAG-2018-01}. Furthermore, the \MET measurement uncertainties are estimated by comparing data to simulation in \Zmm events without jets, as described in \Refn{\cite{ATLAS-CONF-2018-023}}. The uncertainty in the \pileup simulation, derived from a comparison of data with simulation, is also taken into account~\cite{ATLAS-CONF-2014-018}.

The experimental uncertainty related to electron charge misidentification arises from the statistical uncertainty of both the data and the simulated sample of $\Zboson/\gamma^* \rightarrow ee$ events used to measure this probability. The charge misidentification probability increases from $\sim 10^{-4}$ to $\sim 0.1$ with increasing electron \pt and $|\eta|$. The systematic effects obtained by altering the selection requirements imposed on the invariant mass used to select $\Zboson/\gamma^* \rightarrow ee$ events are found to be negligible compared to the statistical uncertainty.
 
The experimental systematic uncertainty in the data-driven estimate of the FNP lepton background is evaluated by varying the nominal fake factor to account for different effects. The $E^{\mathrm{miss}}_{\mathrm{T}}$ requirement is altered to consider variations in the \Wjets composition. The flavour composition of the fakes is investigated by requiring an additional recoiling jet in the electron channel and changing the definition of the recoiling jet in the muon channel. Furthermore, the transverse impact parameter criterion for \textit{tight} muons (defined in \Sect{\ref{sec:ObjSelection}}) is varied by one standard deviation. Finally, in the fake-enriched regions, the normalisation of the subtracted prompt-lepton contribution is altered within its uncertainties. This variation accounts for uncertainties related to the luminosity, the cross-section, and the corrections applied to simulation-based predictions. The FNP systematic uncertainties are correlated across all the different regions and bins, while the statistical uncertainties in each bin are uncorrelated. The statistical uncertainty of the fake factors is added in quadrature to the total systematic uncertainty. The total uncertainty of the FNP background yield varies between \SI{10}{\%} and \SI{20}{\%}.
 
\begin{figure}[htbp]
\centering
\includegraphics[width=\textwidth]{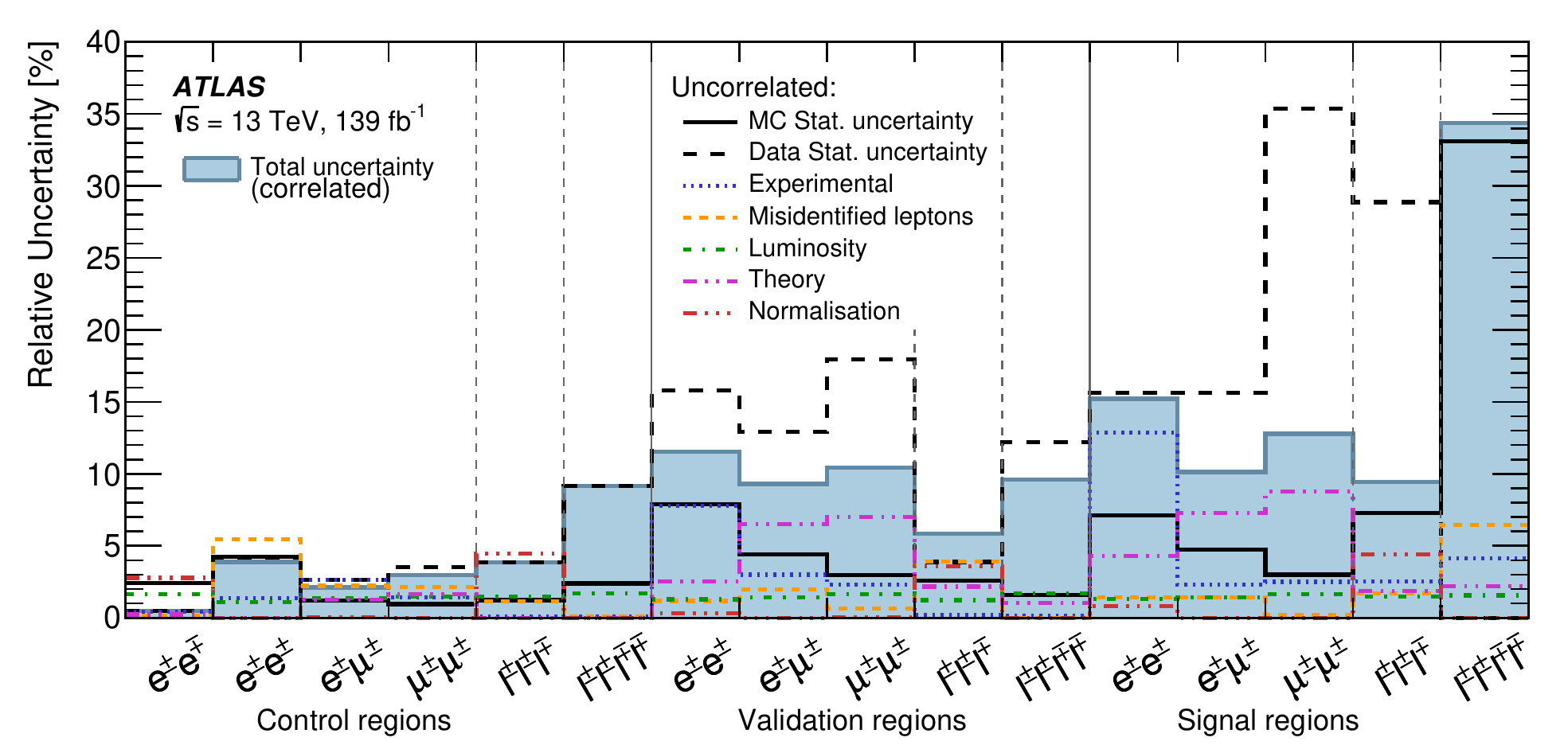}
 
\caption{Relative contributions of different sources of statistical and systematic uncertainty in the total background yield estimation after the background-only fit. Systematic uncertainties are calculated in an uncorrelated way by shifting in turn only one nuisance parameter from the post-fit value by one standard deviation, keeping all the other parameters at their post-fit values, and comparing the resulting event yield with the nominal yield. Validation regions do not constrain the normalisation factors or nuisance parameters. Individual uncertainties can be treated as correlated across the regions. Some backgrounds are constrained by the CR data and have strong anti-correlations as a result. The total background uncertainty is indicated by `Total uncertainty'.}
\label{fig:allAnalysisRegionsSys}
\end{figure}

The total relative background systematic uncertainty after the background-only fit (\Sect{\ref{sec:results}}), and its breakdown into components, is presented in \Fig{\ref{fig:allAnalysisRegionsSys}}.
 
\FloatBarrier

\clearpage

\clearpage
\section{Statistical analysis and results}
\label{sec:results}
 
The \textsc{HistFitter}~\cite{baak2015histfitter} statistical analysis package is used to implement a maximum-likelihood fit. The fit considers the leading lepton pair's invariant mass distribution \mLL, where $\ell, \ell^\prime=e,~\mu$, in all two- and three-lepton control and signal regions. In the four-lepton CRs and SRs, the single-bin event yield is used to obtain the numbers of signal and background events. The binning is chosen to optimise the expected sensitivity to the signal model, while also keeping low statistical uncertainties in each bin. The likelihood is the product of a Poisson probability density function describing the observed number of events and Gaussian distributions to constrain the nuisance parameters associated with the systematic uncertainties. The widths of the Gaussian distributions correspond to the magnitudes of these uncertainties, whereas Poisson distributions are used for MC simulation statistical uncertainties. Furthermore, additional free parameters are introduced for the \DY and diboson background contributions to fit the yields in the corresponding control regions. These free parameters are then used to normalise relevant backgrounds in the signal regions. Fitting the yields of the largest backgrounds reduces the systematic uncertainty in the predicted yield from SM sources. The fitted normalisations are compatible with their SM predictions, within the uncertainties. The diboson yield is described by three free parameters, one each for the two-, three- and four-lepton multiplicity channels. For the scenario in which the \DCH particle decays into the different possible lepton-flavour combinations with equal probability, a \SI{95}{\%} \CL upper limit was set on the $\pp \rightarrow H^{++}H^{--}$ cross-section using the $\text{CL}_{\text{s}}$ method~\cite{read2002presentation}.
 
\subsection{Fit results}
 
\begin{figure}[htbp]
\centering
\includegraphics[width=0.99\textwidth]{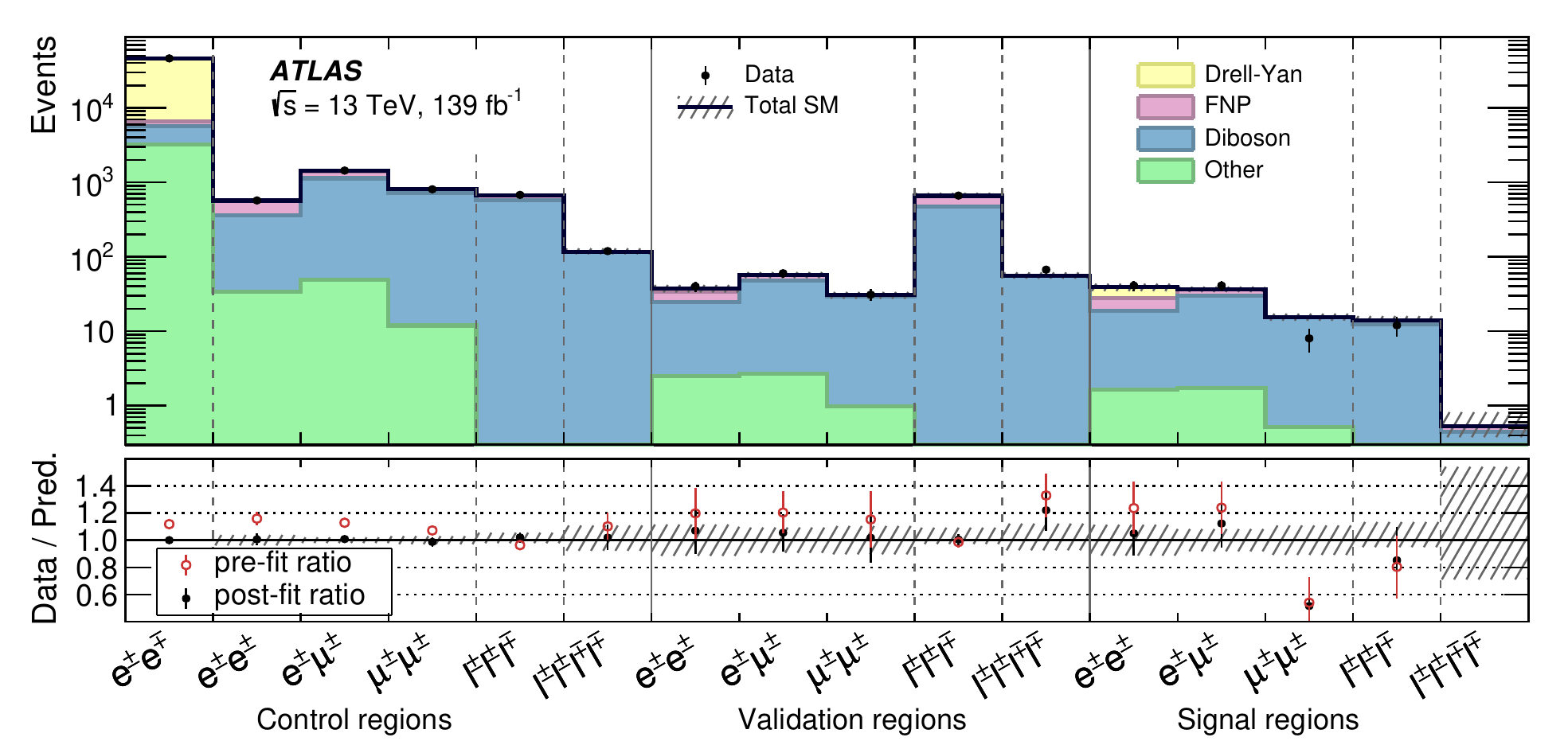}
 
\caption{The numbers of observed and expected events in the control, validation, and signal regions for all channels, split by lepton flavour and electric charge combination. The symbol $\ell$ only stands for light leptons ($\ell, \ell^\prime=e,~\mu$). The background expectation is the result of the background-only fit described in the text. The hatched bands include all post-fit systematic uncertainties with the correlations between various sources taken into account. The error bars show statistical uncertainties. FNP refers to the fake/non-prompt lepton background. Backgrounds from top-quark and multiboson processes are merged, forming the `Other' category. The lower panel shows the ratio of the observed data to the estimated SM background.}
\label{fig:allAnalysisRegions}
\end{figure}

\begin{figure}[t]
\centering
\subfloat[]{
\includegraphics[width=0.45\textwidth]{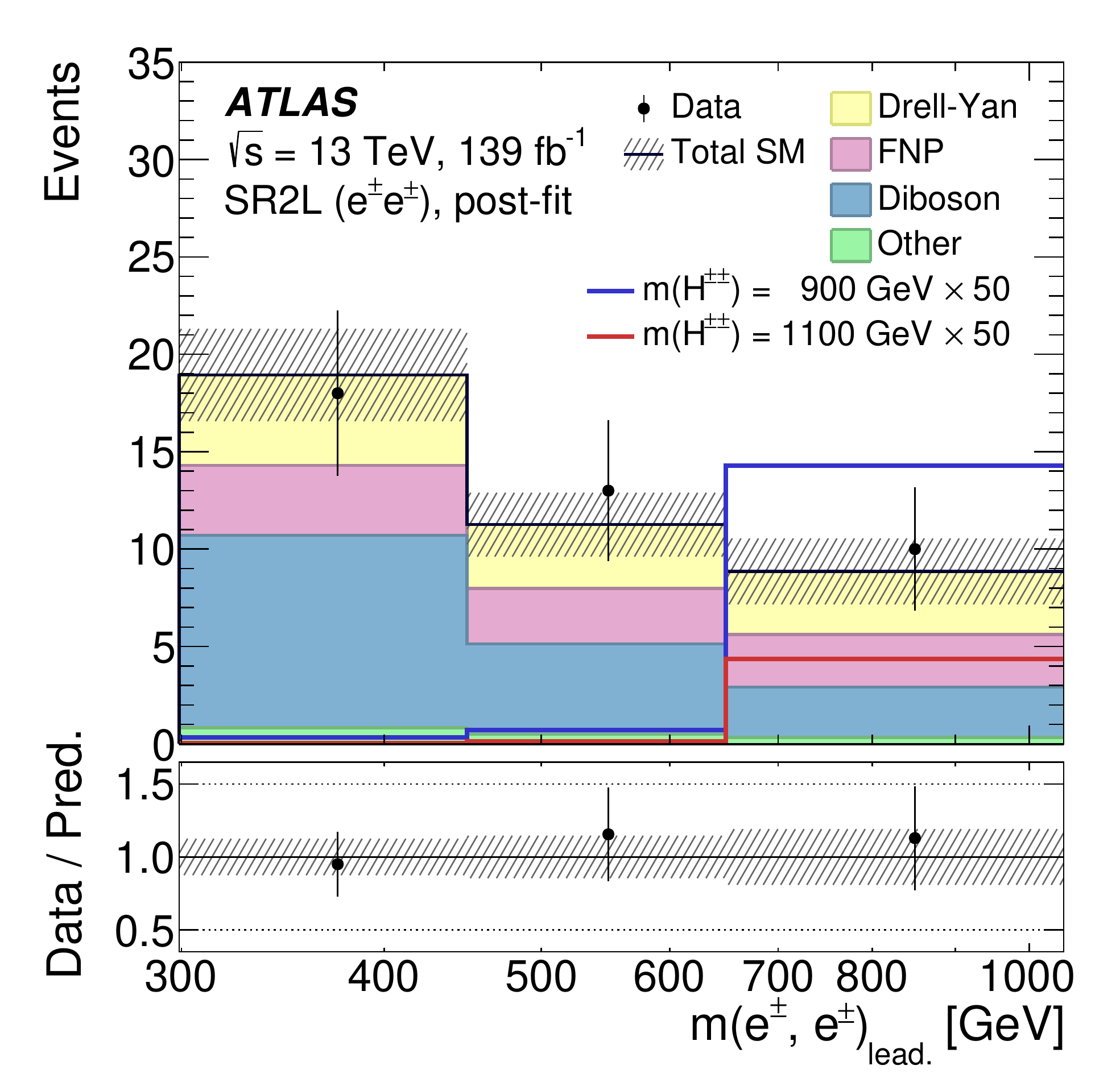}
\label{fig:histo_SR2Lee}
}
\subfloat[]{
\includegraphics[width=0.45\textwidth]{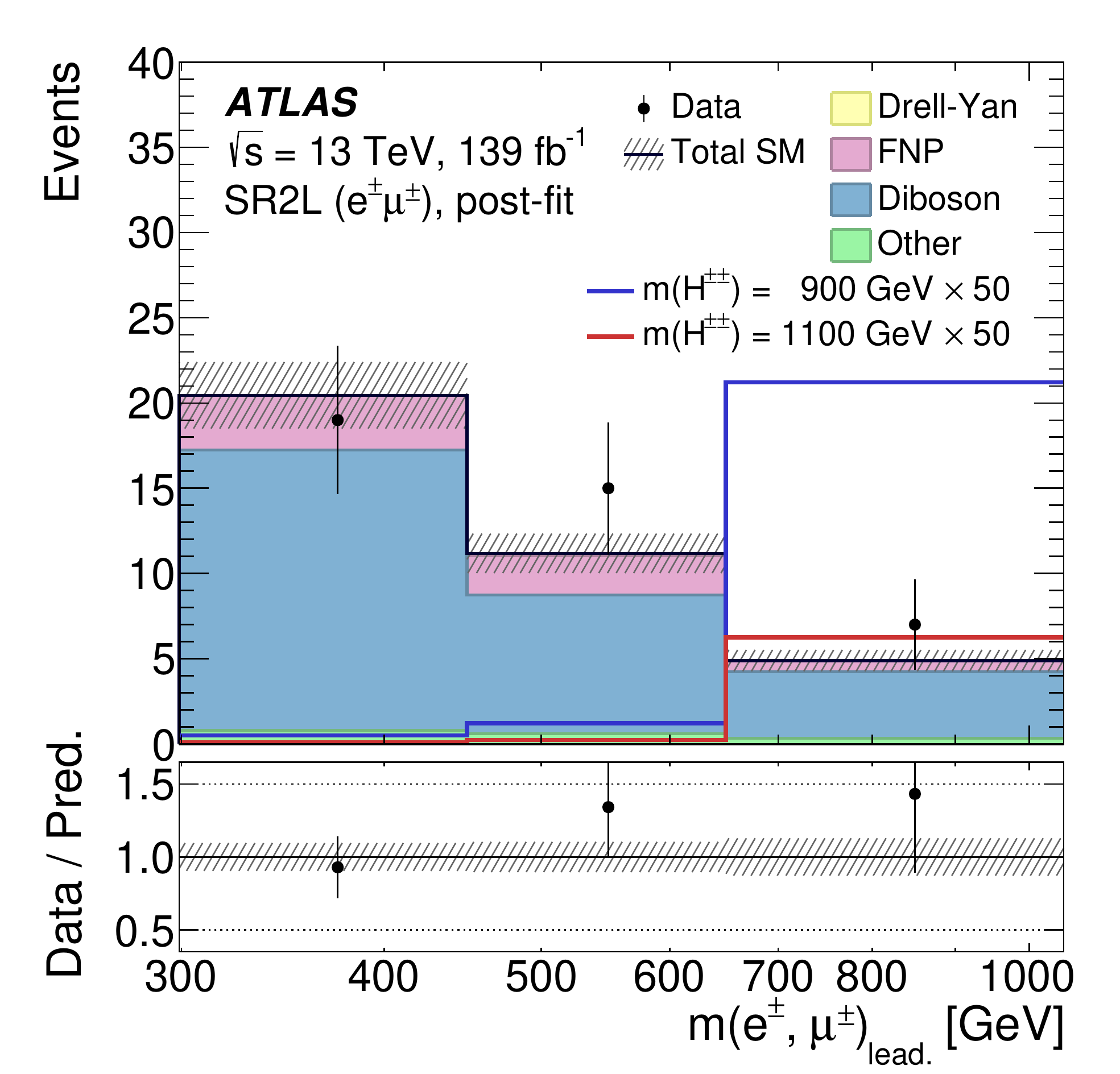}
\label{fig:histo_SR2Lem}
}
\\
\subfloat[]{
\includegraphics[width=0.45\textwidth]{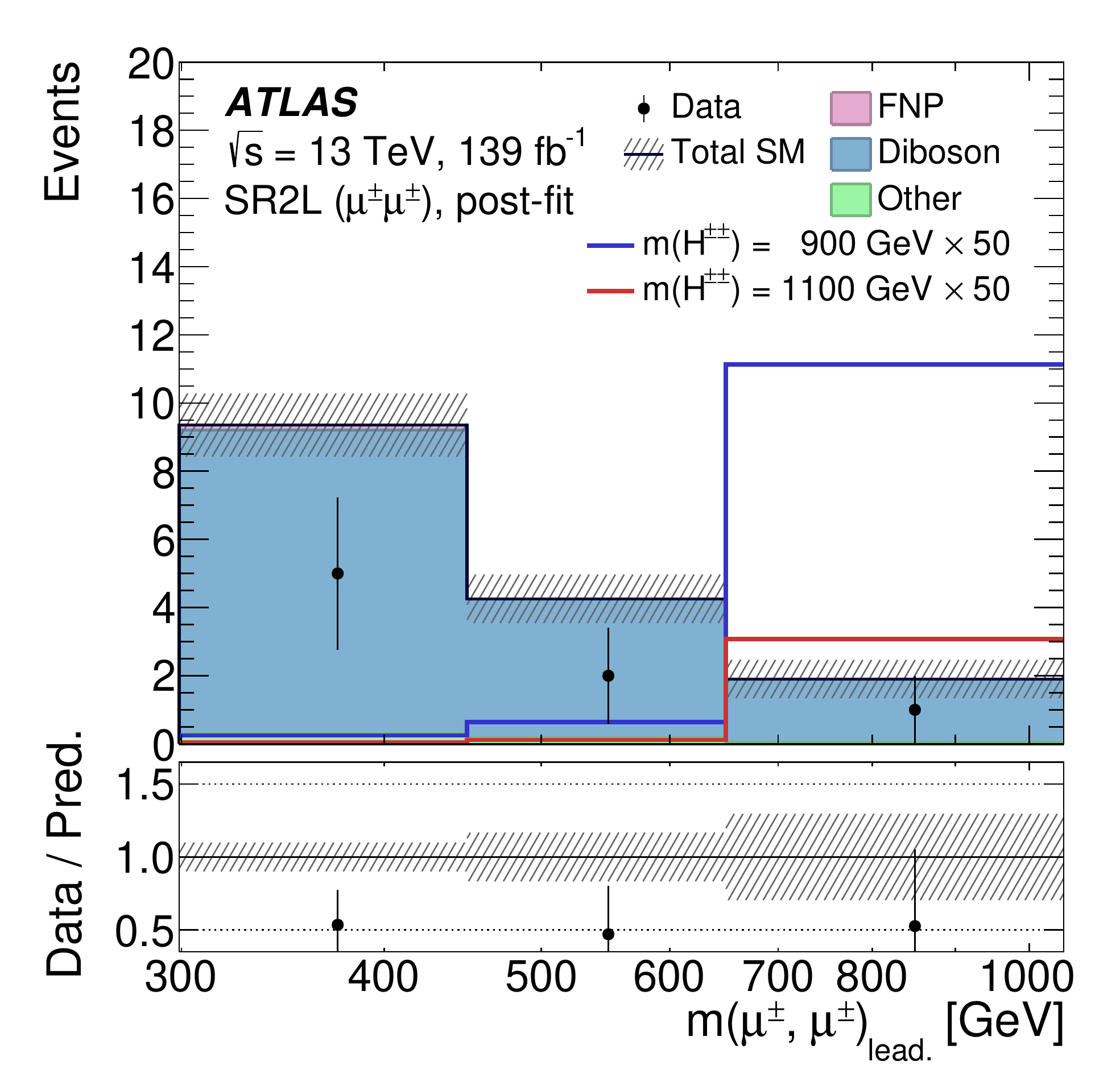}
\label{fig:histo_SR2Lmm}
}
 
\caption{Distributions of \mLL in signal regions, namely
\protect\subref{fig:histo_SR2Lee} the electron--electron two-lepton signal region (SR2L),
\protect\subref{fig:histo_SR2Lem} the electron--muon two-lepton signal region (SR2L) and
\protect\subref{fig:histo_SR2Lmm} the muon--muon two-lepton signal region (SR2L).
The background expectation is the result of the background-only fit described in the text. The hatched bands include all systematic uncertainties post-fit with the correlations between various sources taken into account. The solid coloured lines correspond to signal samples, normalised using the theory cross-section, with the \DCH mass marked in the legend. The $\times 50$ in the legend indicates the scaling of the signal yield to make it clearly visible in the plots. The error bars show statistical uncertainties. Backgrounds from top-quark and multiboson processes are merged, forming the `Other' category. The last bin also includes any overflow entries. The lower panels show the ratio of the observed data to the estimated SM background. The binning presented in the figures is used in the fit.}
\label{fig:representativeSRs}
\end{figure}

\begin{figure}[t]
\centering
\subfloat[]{
\includegraphics[width=0.45\textwidth]{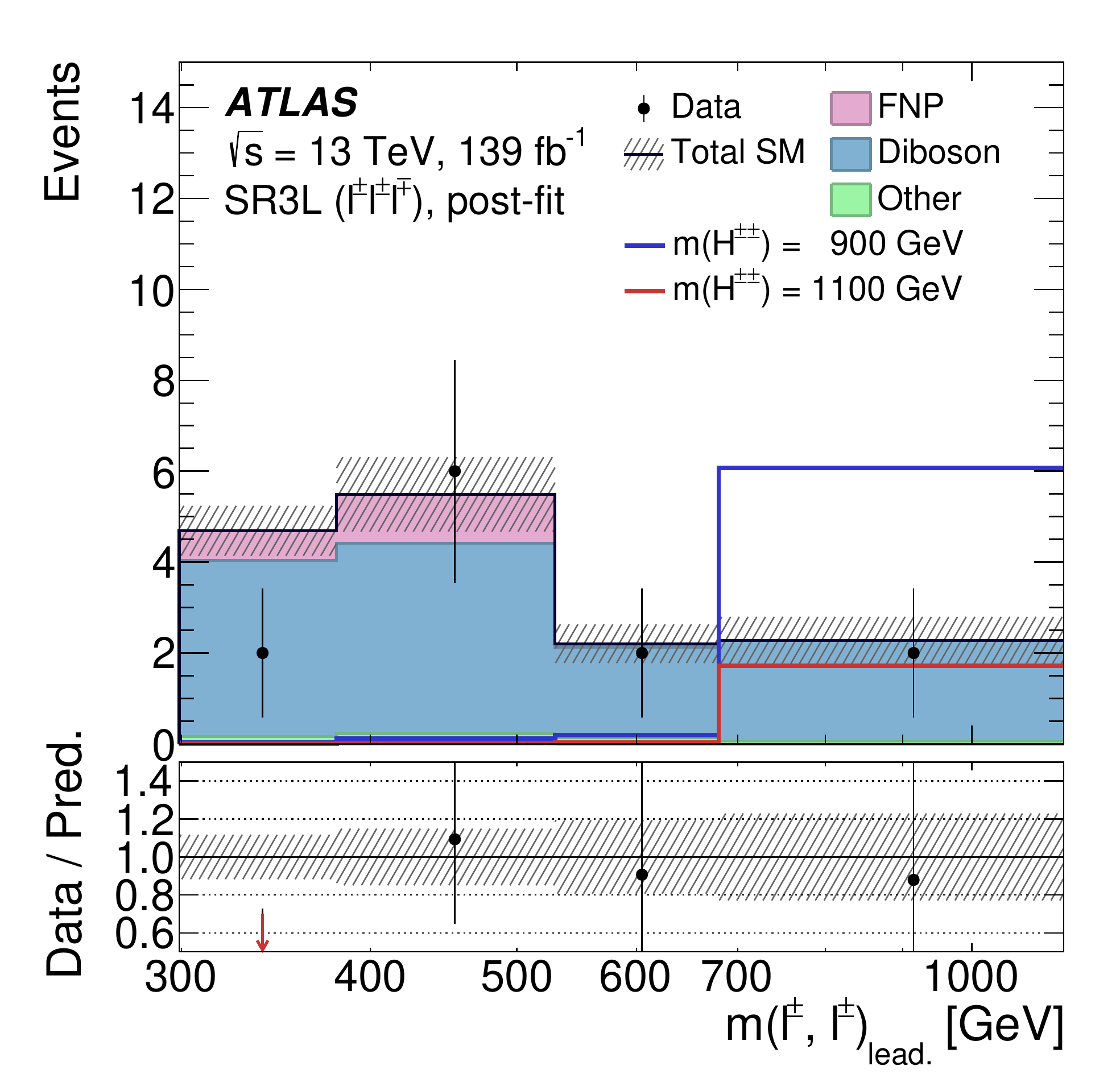}
\label{fig:histo_SR3L}
}
\subfloat[]{
\includegraphics[width=0.45\textwidth]{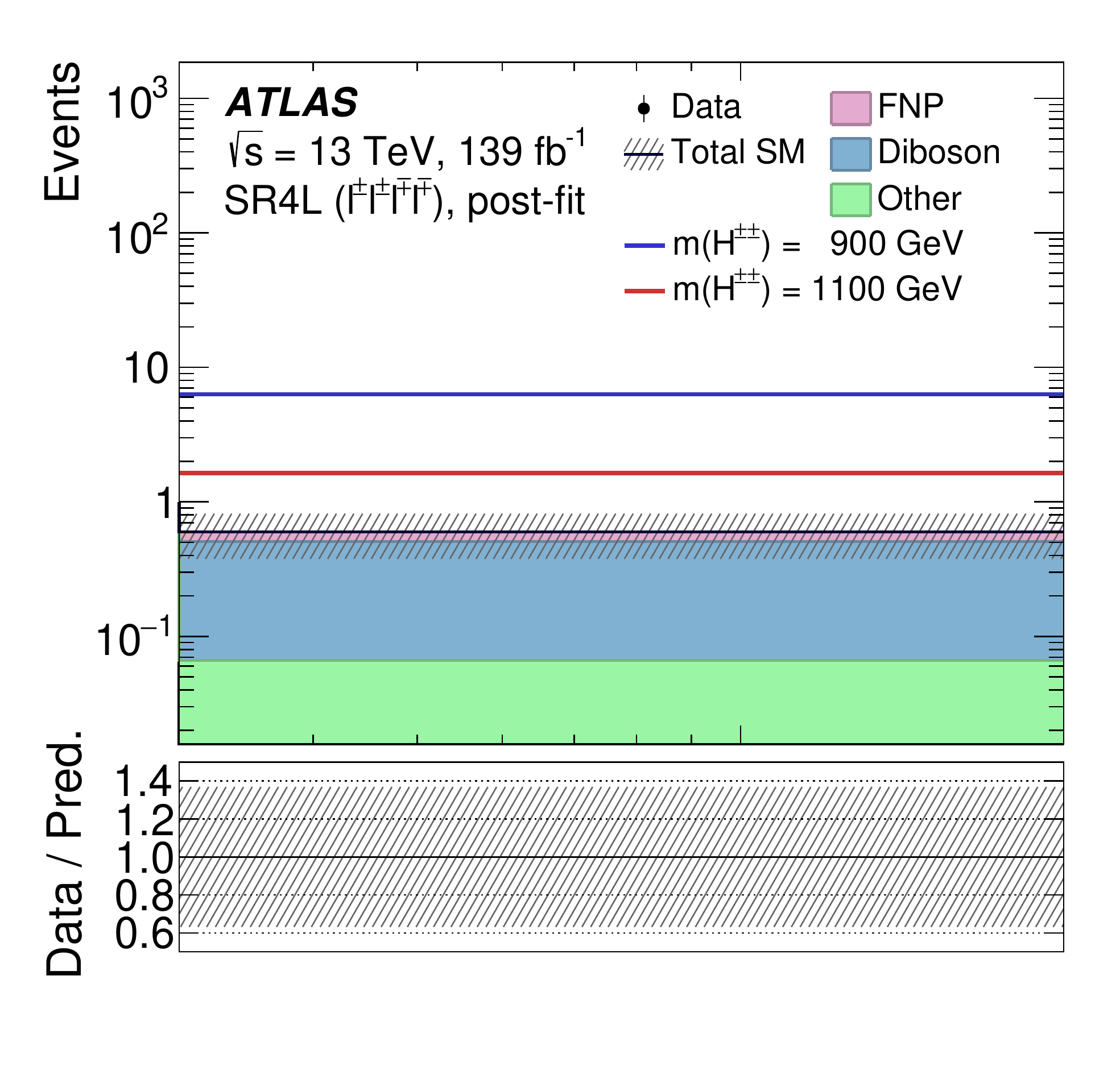}
\label{fig:histo_SR4L}
}
 
\caption{\protect\subref{fig:histo_SR3L} Distribution of \mLL, where $\ell, \ell^\prime=e,~\mu$ in three-lepton signal region (SR3L) and \protect\subref{fig:histo_SR4L} event yield in the four-lepton signal region (SR4L), where no events are observed. The background expectation is the result of the background-only fit described in the text. The hatched bands include all systematic uncertainties post-fit with the correlations between various sources taken into account. The solid coloured lines correspond to signal samples, normalised using the theory cross-section, with the \DCH mass marked in the legend. The error bars show statistical uncertainties. Backgrounds from top-quark and multiboson processes are merged, forming the `Other' category. The lower panels show the ratio of the observed data to the estimated SM background. The binning presented in the figures is used in the fit. \protect\subref{fig:histo_SR3L} The red arrow indicates a point that is outside the vertical range of the figure. The last bin also includes any overflow entries.}
\label{fig:representativeSRs2}
\end{figure}

The observed and expected yields in all control, validation, and signal regions used in the analysis are presented in \Fig{\ref{fig:allAnalysisRegions}} and summarised in \Tabrange{\ref{tab:controlRegions}}{\ref{tab:signalRegions}}. Here `pre-fit' denotes the nominal simulated MC yields and `post-fit' denotes the simulated yields scaled with the normalisation parameters obtained from the likelihood fit to the two-, three-, and four-lepton control and signal regions. In general, good agreement, within statistical and systematic uncertainties, between data and SM predictions is observed in the various regions, demonstrating the validity of the background estimation procedure. No significant excess is observed in any of the signal regions. Correlations between various sources of uncertainty are evaluated and used to estimate the total uncertainty in the SM background prediction. Some of the uncertainties, particularly those connected with the normalisation of the background contributions and the FNP background, are anti-correlated. The \mLL distributions ($\ell, \ell^\prime=e,~\mu$) of the two-lepton signal regions are presented in \Fig{\ref{fig:representativeSRs}}, and those of the three- and four-lepton signal regions are presented in \Fig{\ref{fig:representativeSRs2}}. In the four-lepton signal region, no data event is observed, which is within the expected yield.
 
\begin{figure}[htbp]
\centering
\includegraphics[width=\textwidth]{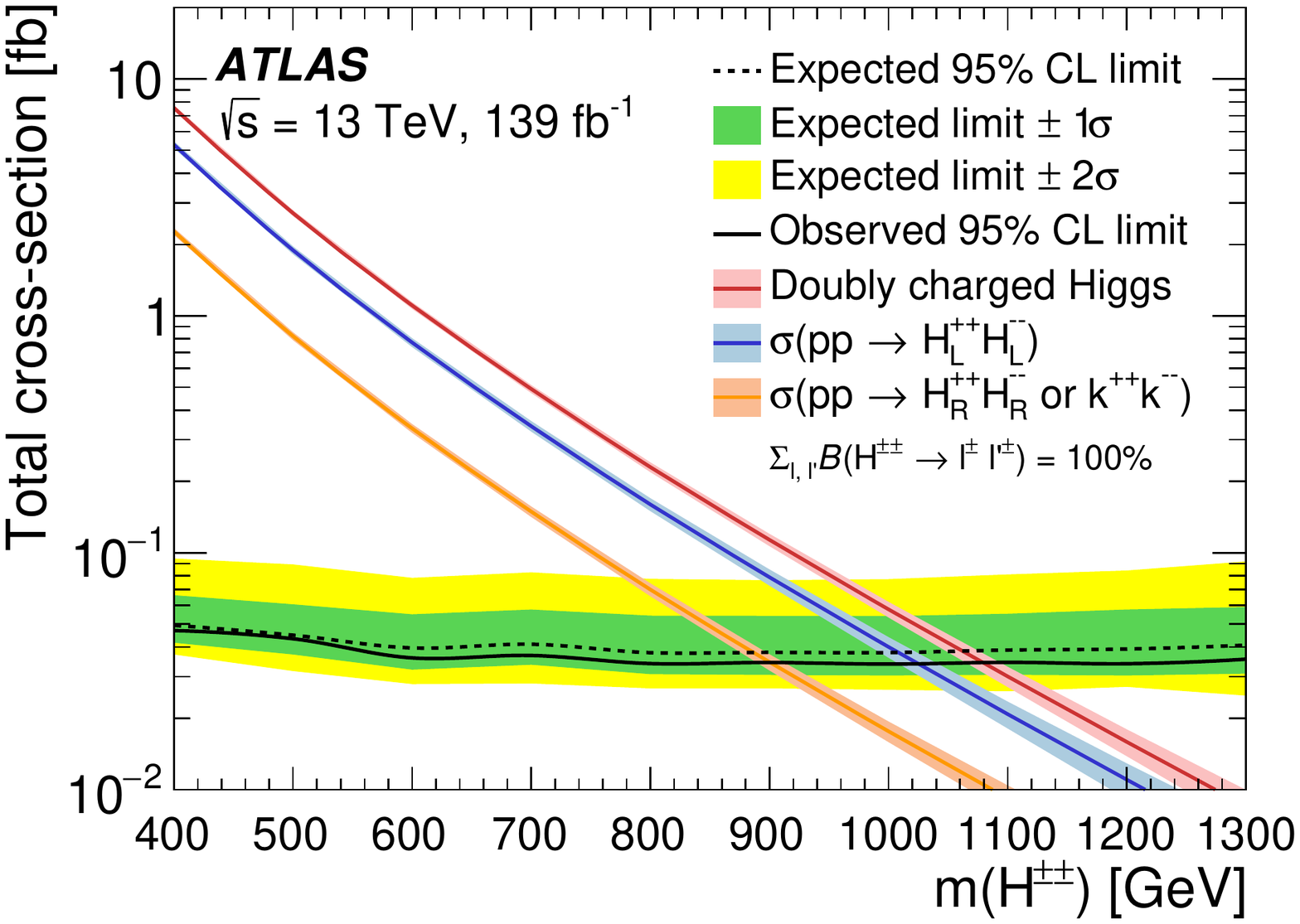}
 
\caption{Observed (solid line) and expected (dashed line) \SI{95}{\%} \CL upper limits on the \DCH pair production cross-section as a function of $m(\DCH)$ resulting from the combination of all analysis channels, assuming $\sum_{\ell \ell^\prime} \BR(\DCH \rightarrow \ell^{\pm} \ell^{\prime \pm})=\SI{100}{\%}$, where $\ell, \ell^\prime = e, \mu, \tau$. The surrounding green and yellow bands correspond to the $\pm 1$ and $\pm 2$ standard deviation ($\pm 1 \sigma$ and $\pm 2 \sigma$) uncertainty around the combined expected limit, respectively, as estimated using the frequentist approach, where toy experiments based on both the background-only and signal+background hypotheses are generated for this purpose. The theoretical signal cross-section predictions, given by the \NLO calculation~\cite{Fuks:2019clu,Ruiz:2022sct}, are shown as blue, orange and red lines for the left-handed $\DCH_L$, right-handed $\DCH_R$ (which is the same as the Zee--Babu $k^{\pm\pm}$), and a sum of the two LRSM chiralities, respectively, with the corresponding uncertainty bands.}
\label{fig:limit_xsec_DCH_full}
\end{figure}

The signal regions were designed to fully exploit the pair production of the \DCH boson and its boosted topology by applying selections that target same-charge high-\pt leptons. After a background-only likelihood fit, the \DY normalisation factor is found to be $1.13 \pm 0.04$ and the two-, three- and four-lepton channel diboson normalisation factors are $1.10 \pm 0.06$, $0.92 \pm 0.05$, and $1.08 \pm 0.11$, respectively. \Fig{\ref{fig:limit_xsec_DCH_full}} shows the upper limit on the cross-section as a function of the \DCH boson mass, where decays into each leptonic final state are equally probable. Since the yields in some regions are very small, the asymptotic approximation~\cite{Cowan:2010js} cannot be used reliably, so $10^5$ pseudo-experiments were run instead to obtain the final limits.
 
\begin{figure}[htbp]
\centering
\includegraphics[width=\textwidth]{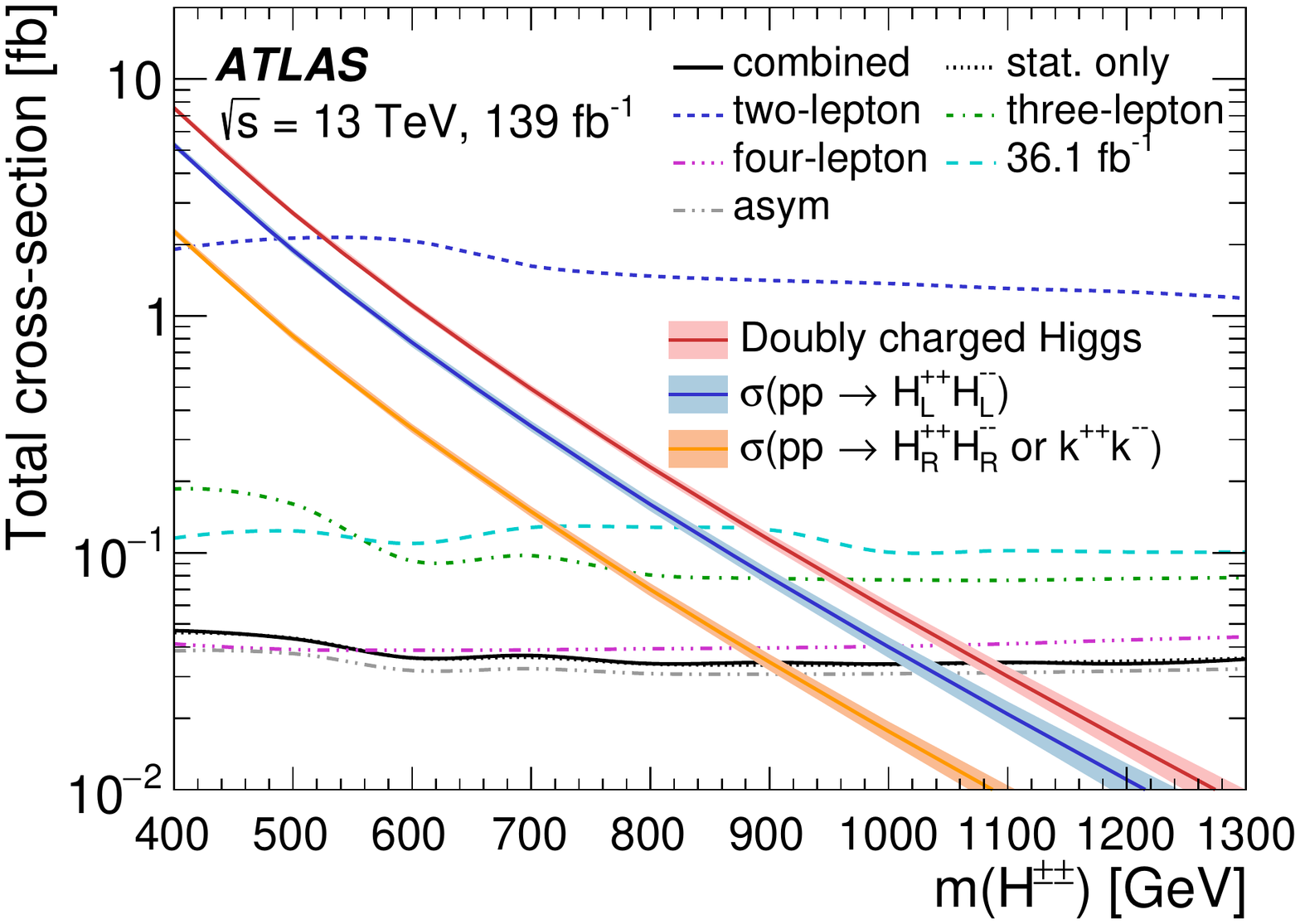}
 
\caption{Observed \SI{95}{\%} \CL upper limits on the \DCH pair production cross-section as a function of $m(\DCH)$ assuming $\sum_{\ell \ell^\prime} \BR(\DCH \rightarrow \ell^{\pm} \ell^{\prime \pm})=\SI{100}{\%}$, where $\ell, \ell^\prime=e, \mu, \tau$. The dashed blue, green, and purple lines indicate the observed limit using the two-, three-, and four-lepton exclusive final states, respectively. The limit obtained from the four-lepton final state is the strongest and drives the combined result. The black lines show the combined observed limit obtained using the frequentist approach for a fit with only statistical uncertainties (dotted) and a fit with statistical and systematic uncertainties (solid). The grey line shows the limit using the asymptotic approximation~\cite{Cowan:2010js}, and the cyan dashed line shows the combined observed limit obtained analysing the first \SI{36.1}{\ifb} of \RunTwo~\cite{EXOT-2016-07}. The theoretical signal cross-section predictions, given by the \NLO calculation~\cite{Fuks:2019clu,Ruiz:2022sct}, are shown as blue, orange and red lines for the left-handed $\DCH_L$, right-handed $\DCH_R$ (which is the same as the Zee--Babu $k^{\pm\pm}$), and a sum of the two LRSM chiralities, respectively, with the corresponding uncertainty bands.}
\label{fig:limit_channels_DCH_full}
\end{figure}

The observed lower limits on the \DCH mass within LRSMs (the Zee--Babu model) vary from \SI{520}{\GeV} to \SI{1050}{\GeV} (\SI{410}{\GeV} to \SI{880}{\GeV}), depending on the lepton multiplicity channel, with $\sum_{\ell \ell^\prime} \BR(H^{\pm \pm}\rightarrow \ell^{\pm}\ell^{\prime \pm}) = \SI{100}{\%}$. The observed lower limit on the mass reaches \SI{1080}{\GeV} and \SI{900}{\GeV} when combining all three channels for LRSMs and the Zee--Babu model, respectively. The expected exclusion limit is \SI[parse-numbers=false]{1065^{+30}_{-50}}{\GeV} for LRSMs and \SI[parse-numbers=false]{880^{+30}_{-40}}{\GeV} for the Zee--Babu model, where the uncertainties of the limit are extracted from the $\pm 1\sigma$ band. The limit obtained from the four-lepton final state is the strongest and drives the combined result. A comparison between the various limits obtained from this measurement is presented in \Fig{\ref{fig:limit_channels_DCH_full}}.
 
\begin{table}[htbp]
\centering
\caption{The number of predicted background events in control regions after the background-only fit, compared with the number of events observed in data. Uncertainties correspond to the uncertainties in the predicted event yields and their total is smaller than the sum of the components in quadrature due to correlations between these components. Due to rounding, the totals can differ from the sums of components. FNP refers to the fake/non-prompt lepton background. Backgrounds from top-quark and multiboson processes are merged, forming the `Other' category. Background processes with a negligible yield are marked with a dash (--).}
\label{tab:controlRegions}

\resizebox{\textwidth}{!}{
\begin{tabular}{l
S[table-format=5.0, round-precision=0] @{$\,\pm\,$}S[table-format=4.0, round-precision=0]
S[table-format=3.0, round-precision=0] @{$\,\pm\,$}S[table-format=2.0, round-precision=0]
S[table-format=4.1] @{$\,\pm\,$}S[table-format=2.1]
S[table-format=3.1] @{$\,\pm\,$}S[table-format=2.1]
S[table-format=3.1] @{$\,\pm\,$}S[table-format=2.1]
S[table-format=3.2] @{$\,\pm\,$}S[table-format=2.2]
}
\toprule
& \multicolumn{2}{c}{DYCR}
& \multicolumn{2}{c}{DBCR2L}
& \multicolumn{2}{c}{DBCR2L}
& \multicolumn{2}{c}{DBCR2L}
& \multicolumn{2}{c}{DBCR3L}
& \multicolumn{2}{c}{CR4L}
\\
& \multicolumn{2}{c}{$e^{\pm} e^{\mp}$}
& \multicolumn{2}{c}{$e^{\pm} e^{\pm}$}
& \multicolumn{2}{c}{$e^{\pm} \mu^{\pm}$}
& \multicolumn{2}{c}{$\mu^{\pm} \mu^{\pm}$}
& \multicolumn{2}{c}{$\ell^{\pm} \ell^{\pm} \ell^{\mp}$}
& \multicolumn{2}{c}{$\ell^{\pm} \ell^{\pm} \ell^{\mp} \ell^{\mp}$}
\\
 
\midrule
 
Observed events
& \multicolumn{2}{c}{\num{46116}}
& \multicolumn{2}{c}{\num{570}}
& \multicolumn{2}{c}{\num{1437}}
& \multicolumn{2}{c}{\num{805}}
& \multicolumn{2}{c}{\num{677}}
& \multicolumn{2}{c}{\num{119}}
\\
 
\midrule
 
Total background
& {\numRF{46116.273842}{4}}      & {\numRF{215.280752}{2}}
& {\numRF{571.054893}{3}}        & {\numRF{23.794098}{2}\pho}
& {\numRF{1440.802082}{2}\phdo}  & {\numRF{37.687638}{1}\phdo}
& {\numRF{797.724545}{3}\phdo}   & {\numRF{27.978805}{2}\phdo}
& {\numRF{674.799898}{3}\phdo}   & {\numRF{25.749708}{2}\phdo}
& {\numRF{118.548059}{3}\phdoo}  & {\numRF{10.863916}{2}\phdoo}
\\
 
\midrule
 
Diboson
& {\pho\numRF{3167.100625}{2}}   & {\numRF{589.6236819}{1}}
& {\numRF{414.214204}{2}}        & {\numRF{73.877596}{1}\pho}
& {\numRF{1168.266054}{3}\phdo}  & {\numRF{71.785431}{1}\phdo}
& {\numRF{686.050254}{3}\phdo}   & {\numRF{32.976794}{2}\phdo}
& {\numRF{595.079284}{3}\phdo}   & {\numRF{30.645141}{2}\phdo}
& {\numRF{116.328629}{3}\phdoo}  & {\numRF{10.873102}{2}\phdoo}
\\
 
FNP lepton
& {\phoo\numRF{663.194325}{2}}     & {\numRF{145.730483}{2}}
& {\numRF{125.254042}{2}}          & {\numRF{65.305016}{1}\pho}
& {\pho\numRF{215.370802}{2}\phdo} & {\numRF{55.834086}{1}\phdo}
& {\numRF{99.806914}{2}\phdo}      & {\numRF{17.404720}{2}\phdo}
& {\pho\numRF{68.817370}{2}\phdo}  & {\numRF{15.344022}{2}\phdo}
& {\phoo\numRF{0.458334}{1}\pho}   & {\pho\numRF{0.367996}{1}\pho}
\\
 
\DY
& {\numRF{39052.104838}{3}}   & {\numRF{554.319911}{1}}
& \multicolumn{2}{c}{--}
& {\phoo\numRF{11.685081}{3}} & {\pho\numRF{2.020811}{2}}
& \multicolumn{2}{c}{--}
& \multicolumn{2}{c}{--}
& \multicolumn{2}{c}{--}
\\
 
Other
& {\pho\numRF{3233.874055}{3}}     & {\numRF{222.140130}{2}}
& {\pho\numRF{31.586647}{2}}       & {\pho\numRF{6.867857}{1}\pho}
& {\phoo\numRF{45.480146}{2}\phdo} & {\pho\numRF{9.233115}{1}\phdo}
& {\pho\numRF{11.867378}{3}}       & {\pho\numRF{0.900216}{1}}
& {\pho\numRF{10.903244}{3}}       & {\pho\numRF{0.644598}{1}}
& {\phoo\numRF{1.761096}{3}}       & {\pho\numRF{0.128815}{2}}
\\
 
\bottomrule
\end{tabular}
}


\end{table}
 
\begin{table}[htbp]
\centering
\caption{The number of predicted background events in validation regions after the background-only fit, compared with the number of events observed in data. Uncertainties correspond to the uncertainties in the predicted event yields and their total is smaller than the sum of the components in quadrature due to correlations between these components. Due to rounding, the totals can differ from the sums of components. FNP refers to the fake/non-prompt lepton background. Backgrounds from top-quark and multiboson processes are merged, forming the `Other' category. Background processes with a negligible yield are marked with a dash (--).}

\small
\begin{tabular}{l
S[table-format=4.0, round-precision=0] @{$\,\pm\,$}S[table-format=3.0, round-precision=0]
S[table-format=4.0, round-precision=0] @{$\,\pm\,$}S[table-format=3.0, round-precision=0]
S[table-format=4.1] @{$\,\pm\,$}S[table-format=2.1]
S[table-format=3.1] @{$\,\pm\,$}S[table-format=2.1]
S[table-format=2.2] @{$\,\pm\,$}S[table-format=1.2]
}
\toprule
& \multicolumn{2}{c}{VR2L}
& \multicolumn{2}{c}{VR2L}
& \multicolumn{2}{c}{VR2L}
& \multicolumn{2}{c}{VR3L}
& \multicolumn{2}{c}{VR4L}
\\
& \multicolumn{2}{c}{$e^{\pm} e^{\pm}$}
& \multicolumn{2}{c}{$e^{\pm} \mu^{\pm}$}
& \multicolumn{2}{c}{$\mu^{\pm} \mu^{\pm}$}
& \multicolumn{2}{c}{$\ell^{\pm} \ell^{\pm} \ell^{\mp}$}
& \multicolumn{2}{c}{$\ell^{\pm} \ell^{\pm} \ell^{\mp} \ell^{\mp}$}
\\
 
\midrule
 
Observed events
& \multicolumn{2}{c}{\num{40}}
& \multicolumn{2}{c}{\num{60}}
& \multicolumn{2}{c}{\num{31}}
& \multicolumn{2}{c}{\num{662}}
& \multicolumn{2}{c}{\num{67}}
\\
 
\midrule
 
Total background
& {\numRF{37.382335}{2}\phdo}      & {\numRF{4.314000}{1}\phdo}
& {\numRF{56.887138}{2}\phdo}      & {\numRF{5.293061}{1}\phdo}
& {\numRF{30.459553}{3}\phoo} & {\numRF{3.175334}{2}\phoo}
& {\numRF{679.934568}{2}\phdoo} & {\numRF{39.663406}{1}\phdoo}
& {\numRF{55.954434}{2}\phdoo}  & {\numRF{5.394421}{1}\phdoo}
\\
 
\midrule
 
Diboson
& {\numRF{22.046833}{3}}   & {\numRF{2.516124}{2}}
& {\numRF{45.477729}{2}\phdo}      & {\numRF{5.337500}{1}\phdo}
& {\numRF{28.437931}{3}\phoo} & {\numRF{3.180967}{2}\phoo}
& {\numRF{471.717963}{3}\phdoo}  & {\numRF{29.605440}{2}\phdoo}
& {\numRF{54.653128}{2}\phdoo}  & {\numRF{5.391216}{1}\phdoo}
\\
 
FNP lepton
& {\pho\numRF{9.227666}{2}}      & {\numRF{0.849581}{1}}
& {\pho\numRF{8.746842}{2}}      & {\numRF{1.278911}{2}}
& {\pho\numRF{0.794387}{2}\pho} & {\numRF{0.200779}{2}\pho}
& {\numRF{186.323369}{3}\phdoo}     & {\numRF{28.557571}{2}\phdoo}
& \multicolumn{2}{c}{--}
\\
 
\DY
& {\pho\numRF{3.612457}{2}} & {\numRF{2.250466}{2}}
& \multicolumn{2}{c}{--}
& {\pho\numRF{0.250585}{3}} & {\numRF{0.027690}{2}}
& {\phoo\numRF{0.922517}{2}}  & {\pho\numRF{0.289221}{2}}
& \multicolumn{2}{c}{--}
\\
 
Other
& {\pho\numRF{2.495379}{2}} & {\numRF{0.572918}{1}}
& {\pho\numRF{2.662567}{2}} & {\numRF{0.519718}{1}}
& {\pho\numRF{0.976649}{2}\pho}  & {\numRF{0.086540}{1}\pho}
& {\pho\numRF{20.970718}{3}\pho}  & {\pho\numRF{1.460113}{2}\pho}
& {\pho\numRF{1.300306}{3}}   & {\numRF{0.086484}{1}}
\\
 
\bottomrule
\end{tabular}


\end{table}
 
\begin{table}[htbp]
\centering
\caption{The number of predicted background events in signal regions after the background-only fit, compared with the number of events observed in data. Uncertainties correspond to the uncertainties in the predicted event yields and their total is smaller than the sum of the components in quadrature due to correlations between these components. Due to rounding, the totals can differ from the sums of components. FNP refers to the fake/non-prompt lepton background. Backgrounds from top-quark and multiboson processes are merged, forming the `Other' category. Background processes with a negligible yield are marked with a dash (--).}
\label{tab:signalRegions}

\small
\begin{tabular}{l
S[table-format=2.1] @{$\,\pm\,$}S[table-format=1.2]
S[table-format=2.2] @{$\,\pm\,$}S[table-format=1.2]
S[table-format=2.3] @{$\,\pm\,$}S[table-format=1.3]
S[table-format=2.2] @{$\,\pm\,$}S[table-format=1.2]
S[table-format=1.3] @{$\,\pm\,$}S[table-format=1.3]
}
\toprule
& \multicolumn{2}{c}{SR2L}
& \multicolumn{2}{c}{SR2L}
& \multicolumn{2}{c}{SR2L}
& \multicolumn{2}{c}{SR3L}
& \multicolumn{2}{c}{SR4L}
\\
& \multicolumn{2}{c}{$e^{\pm} e^{\pm}$}
& \multicolumn{2}{c}{$e^{\pm} \mu^{\pm}$}
& \multicolumn{2}{c}{$\mu^{\pm} \mu^{\pm}$}
& \multicolumn{2}{c}{$\ell^{\pm} \ell^{\pm} \ell^{\mp}$}
& \multicolumn{2}{c}{$\ell^{\pm} \ell^{\pm} \ell^{\mp} \ell^{\mp}$}
\\
 
\midrule
 
Observed events
& \multicolumn{2}{c}{\num{41}}
& \multicolumn{2}{c}{\num{41}}
& \multicolumn{2}{c}{\num{8}}
& \multicolumn{2}{c}{\num{12}}
& \multicolumn{2}{c}{\num{0}}
\\
 
\midrule
 
Total background
& {\numRF{40.318184}{2}\phdoo} & {\numRF{4.520301}{1}\phdoo}
& {\numRF{37.097367}{3}\pho}   & {\numRF{3.121386}{2}\pho}
& {\numRF{14.847504}{3}\phoo}  & {\numRF{1.599383}{2}\phoo}
& {\numRF{14.550090}{3}\pho}   & {\numRF{1.296759}{2}\pho}
& {\numRF{0.619706}{2}\pho}    & {\numRF{0.229134}{2}\pho}
\\
 
\midrule
 
Diboson
& {\numRF{21.255117}{2}\phdoo} & {\numRF{3.955647}{1}\phdoo}
& {\numRF{30.488560}{3}\pho}   & {\numRF{3.264708}{2}\pho}
& {\numRF{14.173579}{3}\phoo}  & {\numRF{1.591172}{2}\phoo}
& {\numRF{12.727383}{3}\pho}   & {\numRF{1.238581}{2}\pho}
& {\numRF{0.443518}{2}\pho}    & {\numRF{0.165397}{2}\pho}
\\
 
FNP lepton
& {\pho\numRF{7.844696}{2}\pho} & {\numRF{1.076065}{2}\pho}
& {\pho\numRF{4.826463}{2}\pho} & {\numRF{1.083279}{2}\pho}
& {\pho\numRF{0.157522}{3}}     & {\numRF{0.034230}{2}}
& {\pho\numRF{1.242742}{2}\pho} & {\numRF{0.458167}{1}\pho}
& {\numRF{0.109532}{2}\pho}     & {\numRF{0.053880}{1}\pho}
\\
 
\DY
& {\pho\numRF{9.660128}{2}\pho} & {\numRF{3.462502}{2}\pho}
& {\pho\numRF{0.148643}{2}}     & {\numRF{0.067262}{1}}
& \multicolumn{2}{c}{--}
& \multicolumn{2}{c}{--}
& \multicolumn{2}{c}{--}
\\
 
Other
& {\pho\numRF{1.558243}{3}}     & {\numRF{0.285535}{2}}
& {\pho\numRF{1.633701}{3}}     & {\numRF{0.248759}{2}}
& {\pho\numRF{0.516402}{2}\pho} & {\numRF{0.048731}{1}\pho}
& {\pho\numRF{0.579965}{2}}     & {\numRF{0.063859}{1}}
& {\numRF{0.066656}{2}}         & {\numRF{0.024500}{2}}
\\
 
\bottomrule
\end{tabular}


\end{table}

\FloatBarrier

\clearpage

\FloatBarrier

\clearpage
\section{Conclusion}
\label{sec:conclusion}
 
The ATLAS detector at the Large Hadron Collider was used to search for doubly charged Higgs bosons in the same-charge two-lepton invariant mass spectrum, using $e^{\pm}e^{\pm}$, $e^{\pm}\mu^{\pm}$ and $\mu^{\pm}\mu^{\pm}$ final states as well as final states with three or four leptons (electrons and/or muons). The search was performed with $\SI{139}{\ifb}$ of data from proton--proton collisions at $\sqrt{s}=\SI{13}{\TeV}$, recorded during the \RunTwo data-taking period lasting from 2015 to 2018. No significant excess above the Standard Model prediction was found. Lower limits are set on the mass of doubly charged Higgs bosons in the context of the left-right symmetric type-II seesaw and Zee--Babu models. These vary between \SI{520}{\GeV} and \SI{1050}{\GeV} for LRSMs and between \SI{410}{\GeV} and \SI{880}{\GeV} for the Zee--Babu model, depending on the lepton multiplicity channel, assuming that $\sum_{\ell \ell^\prime} \BR(\DCH \rightarrow \ell^{\pm}\ell^{\prime \pm})=\SI{100}{\%}$ and that decays to each of the $ee,~e\mu,~\mu\mu,~e\tau,~\mu\tau,~\tau\tau$ final states are equally probable. The observed combined lower limit on the \DCH mass is \SI{1080}{\GeV} within LRSMs and \SI{900}{\GeV} within the Zee--Babu model. These limits are consistent with the expected limits of \SI[parse-numbers=false]{1065^{+30}_{-50}}{\GeV} and \SI[parse-numbers=false]{880^{+30}_{-40}}{\GeV} for LRSM and Zee--Babu doubly charged Higgs bosons, respectively. The lower limits on the LRSM \DCH masses are \SI{300}{\GeV} higher than those from the previous ATLAS result. Moreover, this search provides the first direct test of the Zee--Babu model at the LHC.

\clearpage

\section*{Acknowledgements}


We thank CERN for the very successful operation of the LHC, as well as the
support staff from our institutions without whom ATLAS could not be
operated efficiently.
 
We acknowledge the support of
ANPCyT, Argentina;
YerPhI, Armenia;
ARC, Australia;
BMWFW and FWF, Austria;
ANAS, Azerbaijan;
CNPq and FAPESP, Brazil;
NSERC, NRC and CFI, Canada;
CERN;
ANID, Chile;
CAS, MOST and NSFC, China;
Minciencias, Colombia;
MEYS CR, Czech Republic;
DNRF and DNSRC, Denmark;
IN2P3-CNRS and CEA-DRF/IRFU, France;
SRNSFG, Georgia;
BMBF, HGF and MPG, Germany;
GSRI, Greece;
RGC and Hong Kong SAR, China;
ISF and Benoziyo Center, Israel;
INFN, Italy;
MEXT and JSPS, Japan;
CNRST, Morocco;
NWO, Netherlands;
RCN, Norway;
MEiN, Poland;
FCT, Portugal;
MNE/IFA, Romania;
MESTD, Serbia;
MSSR, Slovakia;
ARRS and MIZ\v{S}, Slovenia;
DSI/NRF, South Africa;
MICINN, Spain;
SRC and Wallenberg Foundation, Sweden;
SERI, SNSF and Cantons of Bern and Geneva, Switzerland;
MOST, Taiwan;
TENMAK, T\"urkiye;
STFC, United Kingdom;
DOE and NSF, United States of America.
In addition, individual groups and members have received support from
BCKDF, CANARIE, Compute Canada and CRC, Canada;
PRIMUS 21/SCI/017 and UNCE SCI/013, Czech Republic;
COST, ERC, ERDF, Horizon 2020 and Marie Sk{\l}odowska-Curie Actions, European Union;
Investissements d'Avenir Labex, Investissements d'Avenir Idex and ANR, France;
DFG and AvH Foundation, Germany;
Herakleitos, Thales and Aristeia programmes co-financed by EU-ESF and the Greek NSRF, Greece;
BSF-NSF and MINERVA, Israel;
Norwegian Financial Mechanism 2014-2021, Norway;
NCN and NAWA, Poland;
La Caixa Banking Foundation, CERCA Programme Generalitat de Catalunya and PROMETEO and GenT Programmes Generalitat Valenciana, Spain;
G\"{o}ran Gustafssons Stiftelse, Sweden;
The Royal Society and Leverhulme Trust, United Kingdom.
 
The crucial computing support from all WLCG partners is acknowledged gratefully, in particular from CERN, the ATLAS Tier-1 facilities at TRIUMF (Canada), NDGF (Denmark, Norway, Sweden), CC-IN2P3 (France), KIT/GridKA (Germany), INFN-CNAF (Italy), NL-T1 (Netherlands), PIC (Spain), ASGC (Taiwan), RAL (UK) and BNL (USA), the Tier-2 facilities worldwide and large non-WLCG resource providers. Major contributors of computing resources are listed in Ref.~\cite{ATL-SOFT-PUB-2021-003}.


\clearpage
 
\printbibliography[notkeyword=aux]
 
\clearpage
 
\begin{flushleft}
\hypersetup{urlcolor=black}
{\Large The ATLAS Collaboration}

\bigskip

\AtlasOrcid[0000-0002-6665-4934]{G.~Aad}$^\textrm{\scriptsize 101}$,
\AtlasOrcid[0000-0002-5888-2734]{B.~Abbott}$^\textrm{\scriptsize 119}$,
\AtlasOrcid[0000-0002-7248-3203]{D.C.~Abbott}$^\textrm{\scriptsize 102}$,
\AtlasOrcid[0000-0002-1002-1652]{K.~Abeling}$^\textrm{\scriptsize 55}$,
\AtlasOrcid[0000-0002-8496-9294]{S.H.~Abidi}$^\textrm{\scriptsize 29}$,
\AtlasOrcid[0000-0002-9987-2292]{A.~Aboulhorma}$^\textrm{\scriptsize 35e}$,
\AtlasOrcid[0000-0001-5329-6640]{H.~Abramowicz}$^\textrm{\scriptsize 150}$,
\AtlasOrcid[0000-0002-1599-2896]{H.~Abreu}$^\textrm{\scriptsize 149}$,
\AtlasOrcid[0000-0003-0403-3697]{Y.~Abulaiti}$^\textrm{\scriptsize 116}$,
\AtlasOrcid[0000-0003-0762-7204]{A.C.~Abusleme~Hoffman}$^\textrm{\scriptsize 136a}$
% ...
\bigskip
\\

% ...
$^{29}$Physics Department, Brookhaven National Laboratory, Upton NY; United States of America.\\
% ...
$^{35}$$^{(a)}$Facult\'e des Sciences Ain Chock, R\'eseau Universitaire de Physique des Hautes Energies - Universit\'e Hassan II, Casablanca;$^{(b)}$Facult\'{e} des Sciences, Universit\'{e} Ibn-Tofail, K\'{e}nitra;$^{(c)}$Facult\'e des Sciences Semlalia, Universit\'e Cadi Ayyad, LPHEA-Marrakech;$^{(d)}$LPMR, Facult\'e des Sciences, Universit\'e Mohamed Premier, Oujda;$^{(e)}$Facult\'e des sciences, Universit\'e Mohammed V, Rabat;$^{(f)}$Institute of Applied Physics, Mohammed VI Polytechnic University, Ben Guerir; Morocco.\\
% ...
$^{55}$II. Physikalisches Institut, Georg-August-Universit\"{a}t G\"ottingen, G\"ottingen; Germany.\\
% ...
$^{101}$CPPM, Aix-Marseille Universit\'e, CNRS/IN2P3, Marseille; France.\\
$^{102}$Department of Physics, University of Massachusetts, Amherst MA; United States of America.\\
% ...
$^{116}$Department of Physics, New York University, New York NY; United States of America.\\
% ...
$^{119}$Homer L. Dodge Department of Physics and Astronomy, University of Oklahoma, Norman OK; United States of America.\\
% ...
$^{136}$$^{(a)}$Departamento de F\'isica, Pontificia Universidad Cat\'olica de Chile, Santiago;$^{(b)}$Millennium Institute for Subatomic physics at high energy frontier (SAPHIR), Santiago;$^{(c)}$Instituto de Investigaci\'on Multidisciplinario en Ciencia y Tecnolog\'ia, y Departamento de F\'isica, Universidad de La Serena;$^{(d)}$Universidad Andres Bello, Department of Physics, Santiago;$^{(e)}$Instituto de Alta Investigaci\'on, Universidad de Tarapac\'a, Arica;$^{(f)}$Departamento de F\'isica, Universidad T\'ecnica Federico Santa Mar\'ia, Valpara\'iso; Chile.\\
% ...
$^{149}$Department of Physics, Technion, Israel Institute of Technology, Haifa; Israel.\\
$^{150}$Raymond and Beverly Sackler School of Physics and Astronomy, Tel Aviv University, Tel Aviv; Israel.\\
% ...
$^{a}$ Also Affiliated with an institute covered by a cooperation agreement with CERN.\\
$^{b}$ Also at Borough of Manhattan Community College, City University of New York, New York NY; United States of America.\\
$^{c}$ Also at Bruno Kessler Foundation, Trento; Italy.\\
$^{d}$ Also at Center for High Energy Physics, Peking University; China.\\
$^{e}$ Also at Centro Studi e Ricerche Enrico Fermi; Italy.\\
$^{f}$ Also at CERN, Geneva; Switzerland.\\
% ...

\end{flushleft}


\end{document}